\documentclass[journal,10pt]{IEEEtran}
\usepackage{amsmath}
\usepackage{bm}
\usepackage{calligra}
\usepackage{amsmath}
\usepackage{subfigure}
\usepackage{graphicx,epsfig,balance}
\usepackage{cite}
\usepackage{color}
\usepackage{multirow}
\usepackage{amsfonts,amssymb}
\usepackage{diagbox}
\usepackage{algpseudocode}
\usepackage{booktabs}
\usepackage{threeparttable}
\usepackage{diagbox}
\usepackage{algorithm}
\usepackage{algorithmicx}
\usepackage{footnote}
\usepackage{float}

\makeatletter
\def\ps@IEEEtitlepagestyle{
  \def\@oddfoot{\mycopyrightnotice}
  \def\@evenfoot{}
}
\def\mycopyrightnotice{
  {\footnotesize
  \begin{minipage}{\textwidth}
  \centering
  Copyright~\copyright~2015 IEEE. Personal use of this material is permitted. However, permission to use this material for any other purposes \\
  must be obtained from the IEEE by sending a request to pubs-permissions@ieee.org.
  \end{minipage}
  }
}

\begin{document}

\title{Semi-Supervised Specific Emitter Identification Method Using Metric-Adversarial Training}
\author{Xue Fu, \IEEEmembership{Student Member,~IEEE},
        Yang Peng, \IEEEmembership{Student Member,~IEEE},
        Yuchao Liu,
        Yun Lin, \emph{Member}, \emph{IEEE},\\
        Guan Gui, \IEEEmembership{Senior Member,~IEEE},
        Haris Gacanin, \IEEEmembership{Fellow,~IEEE},
       and Fumiyuki Adachi, \IEEEmembership{Life Fellow,~IEEE}
\thanks{This work was supported by the National Key Research and Development Program of China under Grant No. 2021ZD0113003, and the Postgraduate graduate Research \& Practice Innovation Programs 2020 Grant No. KYCX20\_0723.~\emph{(Corresponding author: Guan Gui.)}}
\IEEEcompsocitemizethanks{
\IEEEcompsocthanksitem Xue Fu, Yang Peng, and Guan Gui are with College of Telecommunications and Information Engineering, Nanjing University of Posts and Telecommunications, Nanjing 210003, China (e-mail: 1020010415@njupt.edu.cn, 2020010210@njupt.edu.cn, guiguan@njupt.edu.cn).
\IEEEcompsocthanksitem Yuchao Liu is with the China Research Institute of Radiowave Propagation, Qingdao 266107, China (e-mail: lyc8541832@163.com).
\IEEEcompsocthanksitem Yun Lin is with the College of Information and Communication Engineering, Harbin Engineering University, Harbin 150000, China (e-mail: linyun@hrbeu.edu.cn).
\IEEEcompsocthanksitem Haris Gacanin is with the Institute for Communication Technologies and Embedded Systems, RWTH Aachen University, Aachen, Germany (e-mail: harisg@ice.rwth-aachen.de).
\IEEEcompsocthanksitem Fumiyuki Adachi is with the International Research Institute of Disaster Science (IRIDeS), Tohoku University, Sendai 980-8577 Japan (e-mail: adachi@ecei.tohoku.ac.jp).
}}

\markboth{IEEE Internet of Things Journal,~Vol.~XX, No.~XX, XXX~2022}{}
\maketitle

\begin{abstract}
Specific emitter identification (SEI) plays an increasingly crucial and potential role in both military and civilian scenarios. It refers to a process to discriminate individual emitters from each other by analyzing extracted characteristics from given radio signals. Deep learning (DL) and deep neural networks (DNNs) can learn the hidden features of data and build the classifier automatically for decision making, which have been widely used in the SEI research. Considering the insufficiently labeled training samples and large unlabeled training samples, semi-supervised learning-based SEI (SS-SEI) methods have been proposed. However, there are few SS-SEI methods focusing on extracting the discriminative and generalized semantic features of radio signals. In this paper, we propose an SS-SEI method using metric-adversarial training (MAT). Specifically, pseudo labels are innovatively introduced into metric learning to enable semi-supervised metric learning (SSML), and an objective function alternatively regularized by SSML and virtual adversarial training (VAT) is designed to extract discriminative and generalized semantic features of radio signals. The proposed MAT-based SS-SEI method is evaluated on an open-source large-scale real-world automatic-dependent surveillance-broadcast (ADS-B) dataset and WiFi dataset and is compared with state-of-the-art methods. The simulation results show that the proposed method achieves better identification performance than existing state-of-the-art methods. Specifically, when the ratio of the number of labeled training samples to the number of all training samples is 10\%, the identification accuracy is 84.80\% under the ADS-B dataset and 80.70\% under the WiFi dataset. Our code can be downloaded from https://github.com/lovelymimola/MAT-based-SS-SEI.
\end{abstract}


\IEEEpeerreviewmaketitle

\section{Introduction}

Specific emitter identification (SEI) refers to a process to discriminate individual emitters from each other by analyzing extracted characteristics of the received radio signals \cite{Polak_2015}. Extracted characteristics also known as the radio frequency fingerprints (RFFs) that are originated from the imperfections of the analog components of emitters. RFFs are unique to each other and hard to be reproduced \cite{Merchant_2018}. SEI method plays an increasingly important role in cognitive radio (CR) networks \cite{Dobre_2015} and wireless network security \cite{Zhang_2019}. Specifically, SEI methods were proposed as a countermeasure against attackers that disguise themselves as primary users and occupy a licensed part of the spectrum and cause a denial-of-service attack for secondary users in CR \cite{Merchant_2018}. SEI methods were developed as an potential approach of authenticating the identity of a transmitting device for security of Internet of Things (IoT) \cite{ZhaoIoT2022,XieTIFS2021,DeleIoT2020a,Liangiot1,Liangiot2}.

SEI methods can be divided into two types that are transient signals-based SEI methods and steady-state signals-based SEI methods. Initially, RFF-based SEI methods were focusing on identification of transient signals \cite{Toonstra_1995,Toonstra_1996}. Transient signals-based SEI methods use the transition from the turn-off to the turn-on of an emitter that is occurred before the transmission of the actual data of a signal. Therefore, a higher sampling rate is required to extract the transient signal due to its short period. Also, the reliability of the phase and amplitude information is a serious challenge in this area \cite{Kennedy_2008}. In additional, channel noise significantly affects transient signals more than steady-state signals when using radio signals for SEI \cite{Medaiyese_2022}. On the other hand, steady-state signals refer to the modulated parts of the signals transmitted by an emitter at a steady power, and nowadays most of RFFs-based SEI methods are based on steady-state signals.

A typical RFFs-based SEI approach operates in two stages in which the first stage is capturing the signals and the second stage is extracting proper characteristics from captured signals and identifying them \cite{Soltanieh_2020}. Most of the existing SEI methods based on RFF only focus on the second stage on the prerequisite that signals have been captured. Feature extraction and identification of signals is considered as an important stage in the RFF-based SEI methods. Deep learning (DL) and deep neural networks (DNNs) can learn the hidden features of data and build the classifier automatically for decision making in many successful applications \cite{ZhouIoT2022,LinTCCN2020a,KatoIEEE2020,KatoWC2020a,HuangIOT2021b,HuangIOT2021a,WangPHY2021,YuIoT2019}. Because of this feature, many SEI methods considered combining time-domain complex baseband signals and DNN \cite{Merchant_2018, Chen_2019, WangY_2021}. To further improve the identification performance, these methods employed signals in transform domain and DNN, such as bispectrum \cite{Ding_2018}, Hilbert-Huang transform \cite{Pan_2019}, constellation \cite{Peng_2022} and so on. Specifically, Chen \emph{et al.} \cite{Chen_2019} used inception-residual neural network to classify large-scale real-world radio aircraft communications addressing and reporting system (ACARS) and  automatic-dependent surveillance-broadcast (ADS-B) signal data with categories of 3,143 and 5,757, respectively. Wang \emph{et al.} \cite{WangY_2021} used complex-valued neural network (CVNN) with compression to classify 7 power amplifiers (PAs). Merchant \emph{et al.} \cite{Merchant_2018} presented a convolutional neural network (CNN) using time-domain complex baseband error signal for 7 ZigBee devices identification. Ding \emph{et al.} \cite{Ding_2018} adopted a CNN to identify 5 universal software radio peripherals (USRPs) using the compressed bispectrum of the received signal. Pan \emph{et al.} \cite{Pan_2019} presented a deep residual network (ResNet) to identify 5 PAs using Hilbert-Huang spectrum of received signal. Peng \emph{et al.} \cite{Peng_2022} proposed a VSG250 identification method based on heat constellation trace figure (HCTF) and DNN.

The success of DL and DNNs often hinges on the availability of a sufficient number of labeled samples, as shown in Table \ref{Tab:conventional DNN-based SEI or SI methods}, where thousands of samples were labeled to train the DNNs\cite{Qi_2022}. However, in practical SEI tasks, annotation of radio signals is quite expensive, resulting in impossibility of training the DNNs adequately. An attractive approach towards mitigating insufficiently labeled radio signals is semi-supervised learning (SSL) which makes full use of the information embedded in both labeled radio signals and unlabeled radio signals to approach similar performance to that of the well-trained counterpart.

\begin{table*}
  \caption{Related works.}
  \centering
  \subtable[Conventional DNN-based SEI or SI methods]{
  \begin{tabular}{|c|c|c|c|c|c|}
  \hline
  {\bf Conventional Methods}&{\bf DNN}&{\bf Emitter Type}&{\bf Sample Format}&{\bf Number of Samples}&{\bf Performance}\\
  \hline
  \multirow{2}{*}{Chen \emph{et al.} \cite{Chen_2019}}&Inception-residual&3,143 ACARS and &\multirow{2}{*}{IQ}&900,000 and&SNR $\textgreater 9$ dB,\\
  &neural network&5,157 ADS-B&& 13,000,000&$P_{cc} \textgreater 92\%$\\
  \hline
  \multirow{2}{*}{Wang \emph{et al.} \cite{WangY_2021}}&\multirow{2}{*}{CVNN}&\multirow{2}{*}{7 PAs}&\multirow{2}{*}{IQ}&more than 32,000&SNR $\textgreater 10$ dB,\\
  &&&&per device& $P_{cc} \textgreater 75\%$\\
  \hline
  \multirow{2}{*}{Merchant \emph{et al.} \cite{Merchant_2018}}&\multirow{2}{*}{CNN}&\multirow{2}{*}{7 ZigBee}&\multirow{2}{*}{IQ with error}&\multirow{2}{*}{1,000 per device}&SNR $\textgreater 10$ dB,\\
  &&&&&$P_{cc} \textgreater 92.70\%$\\
  \hline
  \multirow{2}{*}{Ding \emph{et al.} \cite{Ding_2018}}&\multirow{2}{*}{CNN}&\multirow{2}{*}{5 USRP}&\multirow{2}{*}{Compressed bispectrum}&0 to 30 dB, 300 from one&SNR $\textgreater 10$ dB,\\
  &&&&device at each SNR&$P_{cc} \textgreater 93\%$\\
  \hline
  \multirow{2}{*}{Pan \emph{et al.} \cite{Pan_2019}}&\multirow{2}{*}{Deep ResNet}&\multirow{2}{*}{5 PAs}&\multirow{2}{*}{Hilbert-Huang spectrum}&$\{10, 12, 14, 16, 18, 20, 22,$&SNR $\textgreater 10$ dB,\\
  &&&&$24\}$ dB, 5,000 per SNR& $P_{cc} \textgreater 62\%$\\
  \hline
  \multirow{2}{*}{Peng \emph{et al.} \cite{Peng_2022}}&InceptionV3&\multirow{2}{*}{7 PAs}&Heat constellation&160 to 260 per &SNR $\textgreater 0$ dB,\\
  &ResNet50, Xception&&trace figure&device&$P_{cc} \textgreater 91.07\%$\\
  \hline
  \end{tabular}
  \label{Tab:conventional DNN-based SEI or SI methods}}

  \subtable[Related SS-SEI or SS-SI methods.]{
  \begin{tabular}{|c|c|c|c|c|c|}
  \hline
  {\bf Related Works}&{\bf SS Framework}&{\bf Emitter Type}&{\bf Sample Format}&{\bf Number of Samples}&{\bf Performance}\\
  \hline
  \multirow{3}{*}{Xie \emph{et al.} \cite{Xie_2022}}&\multirow{3}{*}{CNN and VAT}&\multirow{3}{*}{6 USRPs}&Bispectrum&20,000 per emitter&$R_1$ is $10\%$,\\
  &&&distribution&at a specific SNR&SNR $\textgreater 10$ dB,\\
  &&&&&$P_{cc} \textgreater 93.80\%$\\
  \hline
  \multirow{7}{*}{Gong \emph{et al.} \cite{Gong_2019}}&&&\multirow{7}{*}{IQ}&\multirow{7}{*}{10,000 per emitter}&$R_1$ is $10\%$,\\
  && 5 shortwave stations,&&&$P_{cc} \textgreater 96.50\%$,\\
  &TripleGAN combined&5 PAs, &&&$P_{cc} \textgreater 93.50\%$,\\
  &with autoencoder&5 ultra-shortwave stations,&&& $P_{cc} \textgreater 94.90\%$,\\
  &&5 or 4 Wi-Fi devices&&&$P_{cc} \textgreater 96.20\%$\\
  &&&&&and$P_{cc} \textgreater 96.50\%$,\\
  &&&&&respectively\\
  \hline
  \multirow{3}{*}{Wang \emph{et al.}\cite{WangY_2020}}&Convolutional&\multirow{3}{*}{4 modulations}&\multirow{3}{*}{IQ}&20,000 per&$R_1$ is $5\%$,\\
  &autoencoder&&&modulation at a &SNR $\textgreater 0$ dB,\\
  &&&&specific SNR&$P_{cc} \textgreater 97.65\%$\\
  \hline
  \multirow{2}{*}{Tan \emph{et al.} \cite{Tan_2022}}&\multirow{2}{*}{CGAN}&12 emitters&\multirow{2}{*}{Bispectrum}&1,000 per emitter&$R_2$ is $70\%$,\\
  &&constructed by 6 USRPs&&per modulation&$P_{cc} \textgreater 70\%$\\
  \hline
  \multirow{2}{*}{Ren \emph{et al.} \cite{Ren_2022}}&ResNet18 and Meta&\multirow{2}{*}{15 mobiles phones}&Time-frequency&900 slices&$R_2$ is $1\%$,\\
  &Pseudo Labels&&grayscale image&per model phone&$P_{cc} \approx 91.20\%$\\
  \hline
  \multirow{3}{*}{Medaiyese \emph{et al.} \cite{Medaiyese_2022}}&Denoising&UAV and non-UAV&Hilbert-Huang&\multirow{3}{*}{234,500 slices}&$R_2$ is $10\%$,\\
  &autoencoder and&(bluetooth and Wi-Fi)&and wavelet&&$P_{cc} \approx 84.10\%$\\
  &local autoencoder&&packet transform&&\\
  \hline
  \multicolumn{6}{l}{\small $R_1$ denotes the number of labeled training samples to the number of unlabeled training samples ratio.}\\
  \multicolumn{6}{l}{\small $R_2$ denotes the number of labeled training samples to the number of all training samples ratio.}\\
  \end{tabular}
  \label{Tab:Related Work 1}}

  \subtable[Related ML-SEI or ML-SI methods.]{
  \begin{tabular}{|c|c|c|c|c|c|}
  \hline
  {\bf Related Works}&{\bf DNN and Metric Loss}&{\bf Emitter Type}&{\bf Sample Format}&{\bf Number of Samples}&{\bf Performance}\\
  \hline
  \multirow{2}{*}{Dong \emph{et al.}\cite{Dong_2021}}&CNN and&\multirow{2}{*}{11 modulations}&\multirow{2}{*}{IQ}&ranging from 207&SNR $\textgreater 8$ dB,\\
  &Center Loss&&&to 1,248 per modulation&$P_{cc} \textgreater 86.79\%$\\
  \hline
  \multirow{2}{*}{Shen \emph{et al.} \cite{Shen_2022}}&ResNet and&\multirow{2}{*}{10 LoRa}&Channel independent &500 per device (pretraining) and&\multirow{2}{*}{$P_{cc}\textgreater 98\%$}\\
  &Triplet Loss&&spectrogram of preambles&100 per device (retraining)&\\
  \hline
  \multirow{2}{*}{Gong \emph{et al.} \cite{Gong_2022}}&CNN and&\multirow{2}{*}{10 ISM devices}&\multirow{2}{*}{IQ}&\multirow{2}{*}{10 million per device}&SNR $\textgreater 10$ dB, \\
  &Circle Loss&&&&$P_{cc} \textgreater 94\%$\\
  \hline
  \multirow{3}{*}{He \emph{et al.} \cite{He_2021}}&DNN and&11 and 6 ship-&\multirow{3}{*}{61 acoustic features}&\multirow{3}{*}{Not mentioned}&$P_{cc} \textgreater 80\%$\\
  &Triplet Loss&radiated noise&&& and $P_{cc} \textgreater 90\%$,\\
  &&&&&respectively\\
  \hline
  \multirow{4}{*}{Wang \emph{et al.} \cite{Wang_2022IoT}}&CVNN,&90 ADS-B&\multirow{4}{*}{IQ}&200-500 samples per&\\
  &Triplet Loss&in pretraining, && aircraft in pretraining,&$P_{cc} \textgreater 90\%$\\
  &and &30 ADS-B&& 1-20 samples per &in one-shot\\
  &Center Loss &in finetuning&&aircraft in finetuning&\\
  \hline

  \end{tabular}
  \label{Tab:Related Work 2}}
\end{table*}

In this paper, we propose a semi-supervised learning-based SEI (SS-SEI) method using metric-adversarial training (MAT). Specifically, a well designed object functions that is cross-entropy (CE) loss alternatively regularized by semi-supervised metric learning (SSML) or virtual adversarial training (VAT), where the novel SSML is used to extract the discriminative semantic features of radio signals using Euclidean distance or cosine similarity, and VAT is used to extract the generalized semantic features of radio signals. The main contributions of this paper are summarized as follows:

\begin{itemize}
\item We present MAT-based SS-SEI method, where VAT is used to extract the generalized semantic features of radio signals, and SSML is used to extract the discriminative semantic features of radio signals. VAT and SSML are used alternatively as the regularization term of the objective function, which has a better identification performance and faster convergence rate than simultaneous way.
\item We innovatively introduce the pseudo labels into metric learning (ML), which enables the ML to work for both labeled and unlabeled radio signals on semantic feature space. In addition, this trick is metric-agnostic and we verified the effectiveness on center loss and proxy anchor loss.
\item The proposed SS-SEI method is evaluated on an open source large-scale real-world ADS-B dataset and a open source WiFi dataset, and is compared with four latest SS-SEI methods. The simulation results show that the proposed SS-SEI method achieves the state-of-the-art identification performance.
\end{itemize}

\section{Related Work}
\label{sec2}
In this review, we focus on methods closely related to MAT-based SS-SEI method. MAT is a semi-supervised framework suitable for SEI, containing a variety of semi-supervised  principles such as consistency regularization and pseudo-labels. In addition, ML is another important factor of MAT's success. Therefore, related SS-SEI and semi-supervised learning-based signal identification (SS-SI) methods, and related ML-based SEI (ML-SEI) and ML-based signal identification (ML-SI) methods are reviewed in this paper.

\subsection{SS-SEI and SS-SI Methods.}
In our paper, the semi-supervised (SS) framework is divided into consistency regularization-based framework, entropy minimization-based framework, pseudo label-based framework, unsupervised component-based framework such as autoencoder and generative adversarial network (GAN) and the hybrid framework such as FixMatch \cite{Sohn_2020} that is a combination of consistency regularization and pseudo label.

There are many researchers who made effort on SS-SEI or SS-SI methods based on above SS framework. For example, Xie \emph{et al.} \cite{Xie_2022} proposed a SS-SEI method based bispectrum analysis and CNN with VAT  \cite{Miyato_2019} to identify 6 USRPs. Gong \emph{et al.} \cite{Gong_2019} presented a quadruple-structured framework-based SS-SEI method to identify multiple emitters including PAs, shortwave stations, ultra-shortwave stations and Wi-Fi devices, where the framework consisted of an auto-encoder and a Triple-GAN \cite{Li_2017}. Wang \emph{et al.} \cite{WangY_2020} proposed a convolutional autoencoder for SS-SI. Tan \emph{et al.} \cite{Tan_2022} introduced a GAN using bispectrum of the signal as inputs for SS-SEI and analyzed the identification performance on 12 emitters constructed by 6 USRPs with 6 modulated type. Ren \emph{et al.} \cite{Ren_2022} proposed a SS-SEI method based on ResNet18 and meta pseudo labels \cite{Pham_2021}, where by using the time-frequency grayscale image with short-time Fourier transform (STFT) as the input of ResNet18 and the SS-SEI method was evaluated on dataset of $15$ mobile phones. Medaiyese \emph{et al.} \cite{Medaiyese_2022} presented a hierarchical learning framework for unmanned aerial vehicles (UAVs) detection and identification, where a SS-UAVs detection method based on denosing autoencoder and local outlier factor \cite{Breunig_2000} was introduced. The details of above literatures are shown in Table \ref{Tab:Related Work 1}.

In different SS-SEI or SS-SI methods, unlabeled training dataset participates in training process of DNNs in different ways, which brings different performance benefits. Specifically, SS-SEI method \cite{Xie_2022} have strong anti-noise performance due to training with VAT. SS-SI method \cite{Gong_2019, WangY_2020, Medaiyese_2022}  have strong capability to extract key features because of the utilization of autoencoder. However, the discrimination of features merely brought by softmax with cross-entropy loss and reconstruction loss is limited. DML is one of the solutions to improve discrimination of features.

\subsection{ML-SEI and ML-SI methods}
ML aims to train a DNN that makes tighter and clearer decision boundaries. It can be categorized into pair-based ML and proxy-based ML. The pair-based ML is built upon pairwise distances between samples in the sematic space such as triple loss  \cite{Schroff_2015}, circle loss \cite{Sun_2020} and so on. In the proxy-based ML, each sample is encouraged to be close to proxies of the same category and far apart from those of different category such as center loss \cite{Wen_2016} and proxy-anchor loss \cite{Kim_2020}, where the proxies are representative of a subset of training dataset and learned as a part of the network parameters.

There are many researchers who made effort on SS-SEI or SS-SI methods based on ML. For example, SSRCNN \cite{Dong_2021} was a semi-supervised learning-based automatic modulation classification (SS-AMC) method which consists of a neural network and a sophisticated design of loss functions, where the loss function consists of center loss, cross-entropy loss and kullback-Leibler divergence loss. Shen \emph{et al.} \cite{Shen_2022} exploited channel independent spectrograms of preambles as inputs and a lightweight ResNet as RFF extractor to detect the rogue LoRa devices and classify the legitimate LoRa devices, where the RFF extractor was optimized by triplet loss \cite{Schroff_2015}. Gong \emph{et al.} \cite{Gong_2022} used circle loss \cite{Sun_2020} to optimize a CNN feature extractor and further identify the signals emitted from $10$ different ISM devices. He \emph{et al.} \cite{He_2021} proposed triplet loss which was different from the triplet loss of reference \cite{Schroff_2015} and DNN to classify ship-radiated noise. Wang \emph{et al.} \cite{Wang_2022IoT} presented a well-designed objective function composed of triplet loss and center loss for a discriminative feature embedding and further identified aircrafts in few-shot scenarios. The details of above literatures are shown in Table \ref{Tab:Related Work 2}.

Most of the ML-SEI or ML-SI methods utilize a similarity measure in a fully supervised way or directly combine ML with SS framework where the ML works for labeled samples and the SS framework works for unlabeled samples or all training samples such as SSRCNN \cite{Dong_2021}. Intuitively, there are amount of information embedded in unlabeled samples worth being learnt by ML. However, the lack of labels makes these information inaccessible for ML.

The shortcomings of the above related works can be summarized as limited feature discrimination and information inaccessible of unlabeled samples in ML. To solve these problems, the MAT-based SS-SEI method is proposed in this paper.

\section{Signal Model and Problem Formulation}
\label{sec3}
\subsection{Signal Model}
Only one receiver is employed for a SEI application to capture possible radio signal from a certain interested space. $K$ emitters are considered to be activated at a time and it is assumed that the radio signals from each emitters can be captured individually. The received radio signal for the $k$-th emitters can be formulated as
\begin{align}
\label{fun: ADS-B}
r_k(t) = s_k(t)*h_k(t) + n_k(t), k=1,2,\cdots,K
\end{align}
where $r_k(t)$ is the received radio signal, $s_k(t)$ is the transmitted radio signal, $h_k(t)$ stands for the channel impulse response between transmitter and receiver, $n_k(t)$ denotes a additive white Gaussian noise, and $*$ means the convolution operation.

\subsection{Problem Formulation}
Let $\mathcal{R}$ and $\mathcal{Y}$ be the sample space and category space, respectively. ${\bf{r}}_k  \in \mathcal{R}$ represents the input sample, which is a signal sample from one emitter or with IQ format; $y \in \mathcal{Y}$ denotes the real category of the corresponding emitter.

\subsubsection{SEI problem}
Considering a general machine learning-based SEI problem and a training dataset $\mathcal{D}_t=\{({\bf{r}}_k, y_k)\}_{k=1}^{K}$, the goal of the problem is to produce a mapping function $f \in \mathcal{F}: \mathcal{R} \rightarrow \mathcal{Y}$ and its expected error is minimized, i.e.,
\begin{align}
\label{fun: ex}
\mathop{\min}_{f\in \mathcal{F}}\varepsilon_{ex} = \mathop{\min}_{f\in \mathcal{F}}{\mathbb{E}}_{({\bf{r}}, {\rm{y}}) \sim {\bf{P_{\mathcal{R}\times\mathcal{Y}}}}}\mathcal{L}(f(\bf{r}), \rm{y}),
\end{align}
where $\mathcal{L}({f(\bf{r}}), \rm{y})$ stands for the loss that compares the prediction $f(\bf{r})$ to its ground-truth category. The expected error, however, is approximated by
\begin{align}
\label{fun: em}
\mathop{\min}_{f\in \mathcal{F}}\varepsilon_{em} = \mathop{\min}_{f\in \mathcal{F}}{\mathbb{E}}_{({\bf{r}}, {\rm{y}}) \sim {\mathcal{D}_t}}\mathcal{L}(f(\bf{r}), \rm{y}),
\end{align}
because the joint distribution $\bf{P_{\mathcal{R}\times\mathcal{Y}}}$ is unknown. Therefore, the generalization error $\varepsilon = |\varepsilon_{em}- \varepsilon_{ex}|$ must be considered to prevent overfitting. \eqref{fun: em} can be rewritten as
\begin{align}
\label{fun: em2}
&\mathop{\min}_{f\in \mathcal{F}}\varepsilon_{em},{\rm s.t.}~f({\bf{r}}_{i})={\rm{y}}_{i}, \forall({\bf{r}}_{i}, {\rm{y}}_{i}) \in {\mathcal{D}_t}.
\end{align}
More supervised samples contained in $\mathcal{D}_t$ will bring more constraints on $f$ and then it will bring a good generalization.

\subsubsection{Semi-supervised SEI problem}
In the semi-supervised SEI problem, the training dataset is $D_t = \{({\bf{r}}_1, y_1),\cdots, ({\bf{r}}_L, y_L), {\bf{r}}_{L+1},\cdots, {\bf{r}}_N\}$, where there are $L$ labeled training samples and $u = N- L$ unlabeled training samples. For the convenience of discussion, we use $D_l = \{({\bf{r}}_l^i, y_l^i)|i = 1,\cdots, L\}$ to denote a labeled training dataset, and $ D_{ul} = \{ {\bf{r}}_{ul}^j|j = 1,\cdots, N-L\}$ to denote an unlabeled dataset, The relationship between $D_t$ and $D_l$, $D_{ul}$ is $D_t = D_l \cup D_{ul}$. Usually, $N-L \gg L$.

Considering a general machine learning-based SS-SEI problem, the goal of the SS-SEI problem is also to produce a mapping function $f \in \mathcal{F}: \mathcal{R} \rightarrow \mathcal{Y}$ and its expected error \eqref{fun: ex} is minimized. Due to the limitation of labeled training dataset and additional information on the data distribution from unlabeled training dataset, the expected error, however, is approximated by
\begin{align}
\label{fun: em of SSL}
\mathop{\min}_{f\in \mathcal{F}}\varepsilon_{em} = \mathop{\min}_{f\in \mathcal{F}}{\mathbb{E}}_{({\bf{r}}, {\rm{y}}) \sim {\mathcal{D}_l}}\mathcal{L}(f(\bf{r}), \rm{y}) + {\mathbb{E}}_{({\bf{r}}) \sim {\mathcal{D}_{ul}}}\mathcal{L}_{ul}(*),
\end{align}
where ${L}_{ul}(*)$ stands for the loss that takes into account the unlabeled training dataset to have a more accurate prediction. Reconstruct loss and Kullback-Leibler divergence can be used as the ${L}_{ul}(*)$. There is an important prerequisite that the distribution of samples, which the unlabeled training dataset will help elucidate, are relevant for the SEI problem.

\section{The Proposed MAT-Based SS-SEI Method}
\label{sec4}
In this section, we present the framework of MAT-based SS-SEI in the first sub-section and its training procedure in the second-subsection. The first sub-section shows the overview of the framework, and describes in detail the functions and benefits of each component of the proposed MAT for SS-SEI. The second sub-section shows the training procedure of the proposed MAT-based SS-SEI method.

\subsection{The Framework of MAT-Based SS-SEI Method}

\subsubsection{The overview of the framework}

\begin{figure}[htbp]
  \centering
  \includegraphics[width=3.2 in] {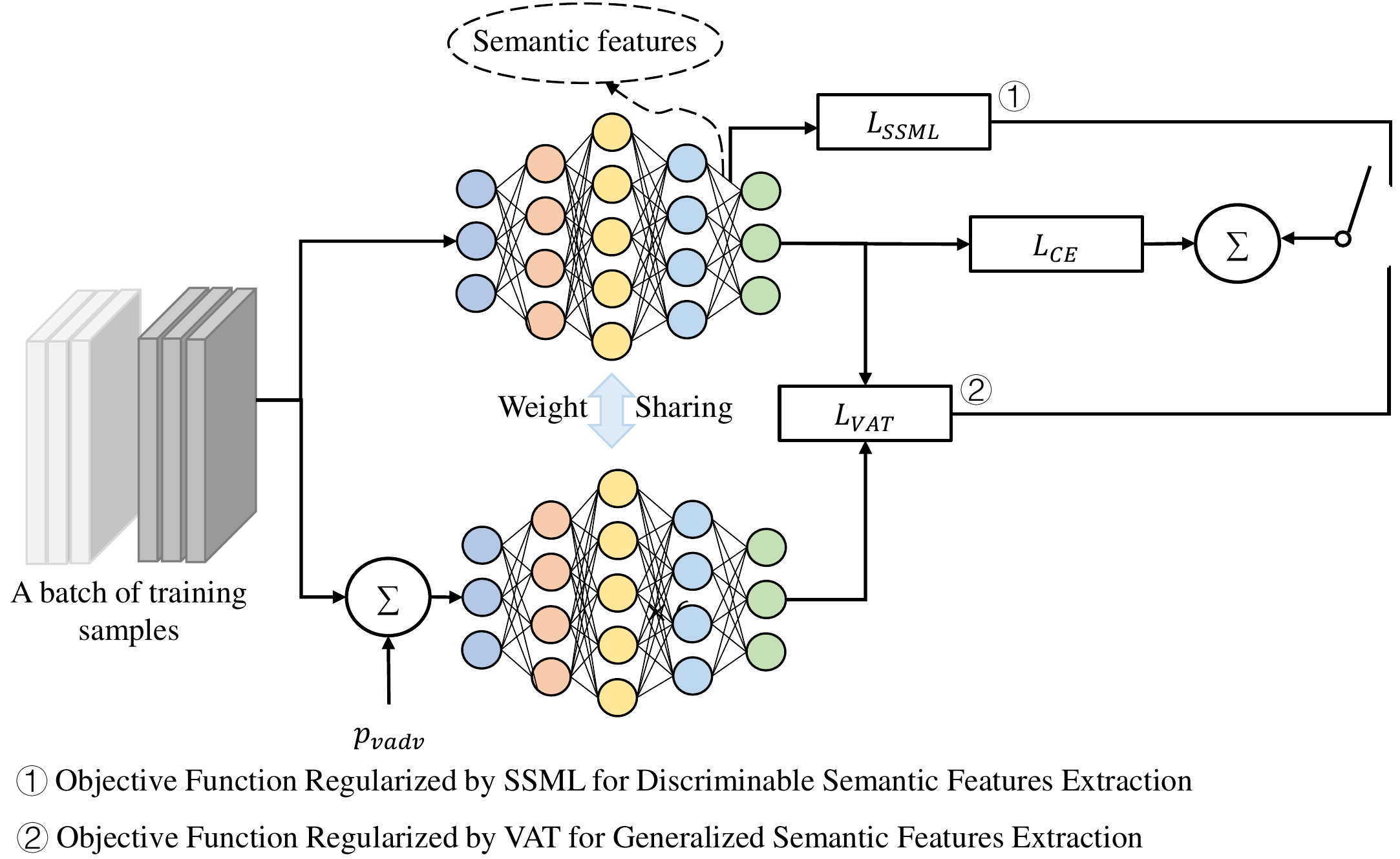}
  \caption{The overview of the proposed MAT-based SEI method.}
  \label{MAT_Pipeline}
\end{figure}

Fig. \ref{MAT_Pipeline} illustrates the overview of our MAT-based SS-SEI method, which contains a DNN and a well-designed objective function alternatively regularized by VAT and SSML.

During the offline training process, firstly, a batch of training samples which consists of labeled samples and unlabeled samples is feeded into the DNN. Secondly, the DNN extracts semantic features and logits of labeled and unlabeled samples. Thirdly, the objective function alternatively regularized by VAT and SSML evaluates the semantic features and logits. Finally, the evaluation results (i.e., the value of objective function) are back-propagated to optimize the DNN. In this paper, CVNN \cite{WangY_2021} is used as the DNN for evaluating the efficacy of our proposed MAT-based SS-SEI method. The structures of CVNN for long signals and short signals are shown as Table \ref{Tab:The structure of CVNN}. It is worthwhile to point that the proposed MAT can be combined with other DNN to achieve other radio signal identification task.

During the online testing process, the testing samples are feed into the optimized DNN (that is CVNN in this paper) and then their category is predicted.

\begin{table}
  \caption{The structure of CVNN.}
  \begin{center}
  \begin{tabular}{|c|c|c|c|}
  \hline
  \multirow{2}{*}{\bf Name}&\multicolumn{2}{|c|}{\bf Structure of CVNN}&\bf Number of\\
  \cline{2-3}
  ~&\bf for long signals&\bf for short signals&\bf layers\\
  \hline
  ~&\multicolumn{2}{|c|}{Complex Conv1D (64,3) + ReLU}&\multirow{2}{*}{$\times 9$}\\
  \multirow{2}{*}{Features}&\multicolumn{2}{|c|}{+ BatchNorm1D + MaxPool1D (2)}&\\
  \cline{2-4}
  \multirow{2}{*}{Extractor}&\multicolumn{2}{|c|}{Flatten}&$\times 1$\\
  \cline{2-4}
  ~&LazyLinear (1024)&LazyLinear (512)&$\times 1$\\
  \cline{2-4}
  ~&/&LazyLinear (128)&$\times 1$\\
  \hline
  Classifier&\multicolumn{2}{|c|}{LazyLinear (K)}&$\times 1$\\
  \hline
  \end{tabular}
  \end{center}
  \label{Tab:The structure of CVNN}
\end{table}

\subsubsection{Semi-supervised classification backbone}
In most of DNN-based SEI methods, the cross-entropy (CE) loss is used as the classification backbone to train the DNN in a fully supervised way. The standard CE loss can be formulated as
\begin{equation}
\begin{split}
\label{fun: CE}
{\mathcal L}_{CE} =  -\frac{1}{L}\sum_{i=1}^{L}\log q_{y_l^i},
\end{split}
\end{equation}
where the ${\bf q}_{l}^i$ is predicted class distribution by the CVNN for ${\bf r}_{l}^i$ and the $q_{y_l^i}$ is the ${y_l^i}$-th value of ${\bf q}_{l}^i$.

In this paper, we introduce semi-supervised cross-entropy (SS-CE) loss to learn the information embedded in unlabeled samples, which can be formulated as
\begin{equation}
\begin{split}
\label{fun: SS-CE}
{\mathcal L}_{CE}^{s} =  -\frac{1}{L}\sum_{i=1}^{L}\log q_{y_l^i} - \frac{1}{N-L}\sum_{j=1}^{N-L} 1\{\mathop{\max}({\bf q}_{ul}^j) \textgreater \tau \} \log q_{\hat y_{ul}^j},
\end{split}
\end{equation}
where $\hat y_{ul}^i$ is the pseudo label of ${\bf r}_{ul}^i$, and ${\bf q}_{ul}^j$ is predicted class distribution by the CVNN for ${\bf r}_{ul}^j$, and $\tau$ is the confidence threshold. The ${\bf q}_{ul}^j$ and $\hat y_{ul}^j$ can be further formulated as
\begin{equation}
\begin{split}
\label{fun: pseudo label1}
{\bf q}_{u l}^j=\operatorname{softmax}\left\{f\left[g\left({\bf r}_{u l}^j\right)\right]\right\},
\end{split}
\end{equation}
\begin{equation}
\begin{split}
\label{fun: pseudo label2}
\hat{y}_{u l}^j=\arg \max _{\mathrm{k}}\left({\bf q}_{ul}^{j}\right),
\end{split}
\end{equation}
where $g(\cdot)$ is the features extractor of CVNN and $f(\cdot)$ is the classifier of CVNN.

\subsubsection{Discriminative semantic features extraction}
Building on the foundation of classification backbone, the CVNN in MAT incorporates ML such that the CVNN is trained to project input samples onto embedding space in which semantically similar features (i.e., radio sigals of the same category) are closely grouped together. Therefore, more discriminative semantic features are extracted compared to CVNN trained only with CE loss. It is worthwhile to point that we innovatively introduce the pseudo labels into ML so that the metric learning can work for both labeled and unlabeled radio signals on semantic feature space and the CVNN in MAT incorporates the semi-supervised metric learning (SSML) for discriminative semantic feature extraction. The proposed SSML is metric-agnostic and the principle of SSML is explained by center loss (CL) \cite{Wen_2016} and proxy-anchor (PA) loss \cite{Kim_2020}.

In this paper, the semantic features are obtained from the features extractor of CVNN, namely the output of LazyLinear (1024) or LazyLinear (128) in Table \ref{Tab:The structure of CVNN}, by which the standard CL can be formulated as
\begin{equation}
\begin{split}
\label{fun: Center}
{\mathcal L}_{Center} =  \frac{1}{2 L}\sum_{i=1}^{L}\left[|{g({\bf r}_l^i)}-{\bf c}_{y_l^i}|^2_2\right],
\end{split}
\end{equation}
where $g({\bf r}_l^i)$ denotes the semantic features of radio signal sample ${\bf r}_l^i$, and ${\bf c}_{y_l^i}$ represents the trainable semantic center features of category $y_l^i$. For each category of radio signals, the CL simultaneously learns a center of semantic features and penalizes the Euclidean distances between its semantic features and corresponding center.
In addition, the standard PA loss can be formulated as
\begin{equation}
\begin{split}
\label{fun: Proxy-anchor}
{\mathcal L}_{PA} =  \frac{1}{|P^+|}\sum_{{\bf{p}} \in P^+}\log\left(1 + \sum_{{\bf{r}}_l \in R_{l}^{p^+}} e ^{- \alpha (s({g({\bf r}_l)}, {\bf{p}})-\delta)}\right)\\
+\frac{1}{|P|}\sum_{{\bf{p}} \in P^-}\log\left(1 + \sum_{{{\bf r}_l} \in R_{l}^{p^-}} e ^{\alpha (s(g({\bf r}_l), {\bf{p}})+\delta)}\right),
\end{split}
\end{equation}
where $\delta > 0$ represents a margin, and $\alpha > 0$ denotes a scaling factor, and $P$ is the set of all proxies, and $P^+$ stands for the set of positive proxies of semantic features, and $P^-$ stands for the set of negative proxies of semantic features, and $s(\cdot, \cdot)$ denotes the cosine similarity between two features, and $R_{l}^{p^+}$ denotes the set of positive labeled samples of ${\bf{p}}$, and $R_{l}^{p^-} = R_{l} - R_{l}^{p^+}$ denotes the set of negative labeled samples of ${\bf{p}}$.

Building on the foundation of standard ML, we introduce the pseudo labels into standard ML. For the CL, the semi-supervised center loss (SS-CL) can be formulated as
\begin{equation}
\begin{split}
\label{fun: SS-Center}
{\mathcal L}_{Center}^{s} =  \frac{1}{2 L}\sum_{i=1}^{L}\left[|{g({\bf r}_l^i)}-{\bf c}_{y_l^i}|^2_2\right] \\
+ \frac{1}{2 (N-L)}\sum_{j=1}^{N-L}1\{\mathop{\max}({\bf q}_{ul}^j) \textgreater \tau \}\left[|{g({\bf r}_{ul}^j)}-{\bf c}_{\hat y_{ul}^j}|^2_2\right],
\end{split}
\end{equation}

In same way, the semi-supervised proxy anchor (SS-PA) loss can be formulated as
\begin{equation}\scriptsize
\begin{split}
\label{fun: SS-Proxy-anchor}
{\mathcal L}_{PA}^{s} =  \frac{1}{|P^+|}\sum_{{\bf{p}} \in P^+}\log\left(1 + \sum_{{\bf{r}}_l \in R_{l}^{p^+}} e ^{- \alpha (s({g({\bf r}_l)}, {\bf{p}})-\delta)}\right)\\
+ \frac{1}{|P|}\sum_{{\bf{p}} \in P^-}\log\left(1 + \sum_{{{\bf r}_l} \in R_{l}^{p^-}} e ^{\alpha (s(g({\bf r}_l), {\bf{p}})+\delta)}\right) \\
+ \frac{1}{|P^+|}\sum_{{\bf{p}} \in P^+}1\{\mathop{\max}({\bf q}_{ul}^j) \textgreater \tau \}\log\left(1 + \sum_{{\bf{r}}_{ul} \in R_{ul}^{p^+}} e ^{- \alpha (s({g({\bf r}_{ul})}, {\bf{p}})-\delta)}\right) \\
+ \frac{1}{|P|}\sum_{{\bf{p}} \in P^-}1\{\mathop{\max}({\bf q}_{ul}^j) \textgreater \tau \}\log\left(1 + \sum_{{{\bf r}_{ul}} \in R_{ul}^{p^-}} e ^{\alpha (s(g({\bf r}_{ul}), {\bf{p}})+\delta)}\right),
\end{split}
\end{equation}

Intuitively, the SS-CE loss forces the semantic features of different categories staying apart roughly. The SS-CL efficiently pulls the semantic features of the same categories to their center using Euclidean distance. The SS-PA loss enlarges the inter-category difference, but also reduces intra-category variations using cosine similarity. With the joint loss function of CE loss and SS-CL or SS-PA loss, the CVNN is trained to obtain the semantic features with inter-category dispersion and intra-category compactness as much as possible. The objective function regularized by SSML for discriminative semantic features extraction can be formulated as
\begin{equation}
\begin{split}
\label{fun: joint loss2}
{\mathcal L}_{2} = {\omega}_3 {\mathcal L}_{CE}^s + {\omega}_4 {\mathcal L}_{SSML},
\end{split}
\end{equation}
where ${\mathcal L}_{SSML}$ is ${\mathcal L}_{Center}^{s}$ or ${\mathcal L}_{PA}^{s}$, and scalars $\omega_3$ and $\omega_4$ are used for balancing the two loss terms. Different scalars lead to different semantic features distributions. With proper scalars, the discrimination of semantic features can be significantly improved. In this paper, we use the automatic weight \cite{liebei_2018} to get rid of the manual tuning of scalars.

\subsubsection{Generalized semantic features extraction}
In practice, the evaluation of the objective function will always be an empirical approximation over the sample space as illustrated in equation \eqref{fun: em}, and however the number of the samples that can be used to tune the parameters of model is finite, especially in SS scenarios. Therefore, even with successful optimization and low training error, the testing error can be large in the SS-SEI, that is, the generalization performance of model are not sufficient. It is known that the generalization performance of DNNs can be improved by applying random perturbations to samples to generate artificial input samples and encouraging the DNNs to assign similar output to the set of artificial input samples derived from the same samples \cite{Bishop_1995}. Adversarial training \cite {Goodfellow_2015} is one of the successful attempts that improve generation performance by applying random perturbations. VAT \cite{Miyato_2019} is an improved adversarial training which can be applied to the SSL, and we use VAT to achieve generalized semantic features extraction in our SS-SEI method.

The VAT defines the local distributional smoothness (LDS) to be the divergence-based distributional robustness of the model against virtual adversarial direction, and LDS can be formulated as
\begin{equation}
\begin{split}
\label{fun: LDS of VAT}
\mathbf{LDS}({\bf{r}}_*, \theta) = D[f(y|{\bf{r}}_*, \hat{\theta}), f(y|{\bf{r}}_* + \bf{p_{vadv}}, \theta)],
\end{split}
\end{equation}
\begin{equation}
\begin{split}
\label{fun: r of VAT}
{\bf{p}}_{vadv} = \arg \mathop{\max}_{\bf{p};||\bf{p}||_2 < \epsilon}D[f(y|{\bf{r}}_*,\hat{\theta}), f(y|{\bf{r}}_* + {\bf{p}}, \theta)],
\end{split}
\end{equation}
where ${\bf{r}}_*$ represents either ${\bf r}_l$ or ${\bf r}_{ul}$, $D[f, f']$ is kullback-Leibler divergence in our SS-SEI method, $\hat{\theta}$ stands for the current model weight, $f(y|{\bf{r}}_*, \hat{\theta})$ is current estimation of true distribution of the output label of ${\bf{r}}_*$, $f(y|{\bf{r}}_* + {\bf{p}}, \theta)$ is the current estimation of distribution of the output label of ${\bf{r}}_*$ with virtual adversarial perturbation, and ${\bf p}_{vadv}$ is virtual adversarial perturbation which can be approximated by
\begin{equation}
\begin{split}
\label{fun: r of VAT approximated}
{\bf{p}}_{vadv} \approx \epsilon \frac{g}{||g||_2},
\end{split}
\end{equation}
where $g = {\nabla}_{{\bf{r}}_*} D[f(y|{\bf{r}}_*,\hat{\theta}), f(y|{\bf{r}}_* + {\bf{p}}, \theta)]$ and $\epsilon$ represents perturbation intensity. The gradient ${\nabla}_{r_*} D[f(y|{\bf{r}}_*,\hat{\theta}), f(y|{\bf{r}}_* + {\bf{p}}, \theta)]$ can be efficiently computed by back-propagation of the CVNN.

The SS-CE loss is used as the classification backbone and the LDS is used as a way for enhancing the generalization performance of model. The objective function regularized by VAT for generalized semantic features extraction can be formulated as
\begin{equation}
\begin{split}
\label{fun: joint loss1}
{\mathcal L}_{1} =  {\omega}_1 {\mathcal L}_{CE}^s + {\omega}_2 {\mathcal L}_{VAT},
\end{split}
\end{equation}
\begin{equation}
\begin{split}
\label{fun: VAT}
{\mathcal L}_{VAT} =  \frac{1}{N_l + N_{ul}}\sum_{{\bf{r}}_* \in D_l,D_{ul}}\mathbf{LDS}({\bf{r}}_*, \theta),
\end{split}
\end{equation}
where scalars $\omega_1$ and $\omega_2$ are used for balancing the two loss terms. In this paper, we use the automatic weight \cite{liebei_2018} to get rid of the manual tuning of scalars.

\subsection{Training Procedure}
The full Training procedure with object function is described in Algorithm \ref{alg:Training Procedure}. Alternative optimization is used during training. Specifically, the objective function regularized by VAT for generalized semantic features extraction, that is \eqref{fun: joint loss1}, is operated when $t \in \{1, 3, 5,\cdots, T-1\}$, and the objective function regularized by SSML for discriminative semantic features extraction, that is \eqref{fun: joint loss2}, is operated when $t \in \{2, 4, 6,\cdots, T\}$.

\begin{algorithm}[htb]
\small
\caption{Training Procedure of MAT-based SS-SEI.}
\label{alg:Training Procedure}
\textbf {Require}:
\begin{itemize}
  \item $D_l$, $D_{ul}$: Labeled and unlabeled training dataset, respectively;
  \item $T$: Number of training iterations;
  \item $B$: Number of batches in a training iteration;
  \item $\theta_m$, $\theta_a$: Parameters of CVNN, trainable semantic center features of center loss or proxy-anchor loss, respectively;
  \item $lr_m$, $lr_a$: Learning rate of CVCNN, center loss or proxy-anchor loss, respectively;
  \item $\omega_1$, $\omega_2$, $\omega_3$, $\omega_4$: Scalars for balancing the loss terms;
\end{itemize}

\begin{algorithmic}[1]
\Statex {\bf Dataset preprocessing}:
    \State $D_l \leftarrow \frac{D_l-\min(D_l\cup D_{ul})}{\max(D_l\cup D_{ul})-\min(D_l\cup D_{ul})}$;
    \For {$t=1$ to $T$}:
        \For {$b=1$ to $B$}:
           \State Sample a batch of labeled training samples $({\bf{r}}_l, y_l)$.
           \State Sample a batch unlabeled training samples $({\bf{r}}_{ul})$.
           \If {$t\%2 == 0$}:
           \Statex \quad\quad\quad\quad{\bf Forward propagation}:
                \State ${\bf z}_l, {\bf z}_{ul} = g({\bf \theta}^{t,b}_m,{\bf \theta}^{t,b}_a; {\bf r}_l, {\bf r}_{ul})$;
                \State ${\bf \hat{p}}_l, {\bf \hat{p}}_{ul} = f({\bf \theta}^{t,b}_m; {\bf z}_l, {\bf z}_{ul})$;
                \State $\mathcal L_{2} = \omega_3 \mathcal L_{CE}^s(({\bf \hat{p}}_l, {\bf \hat{p}}_{ul}), ({y}_l,{\hat y}_{ul})) +$
            \Statex \quad\quad\quad\quad\quad\quad\quad $\omega_4 \mathcal L_{SSML}(({\bf z}_l, {\bf z}_{ul}), {\bf \theta}_a)$;
            \Statex \quad\quad\quad\quad{\bf Backward propagation}:
                \State ${{\bf \theta}^{t,b+1}_m} \leftarrow Adam({\nabla}_{\theta_m}, \mathcal L_{2}, lr_m, \theta_m)$
                \State ${{\bf \theta}^{t,b+1}_a} \leftarrow Adam({\nabla}_{\theta_a}, \mathcal L_{2}, lr_a, \theta_a)$
            \Else:
            \Statex \quad\quad\quad\quad{\bf Forward propagation}:
                \State ${\bf z}_l, {\bf z}_{ul} = g({\bf \theta}^{t,b}_m,{\bf \theta}^{t,b}_a; {\bf r}_l, {\bf r}_{ul})$;
                \State ${\bf \hat{p}}_l, {\bf \hat{p}}_{ul} = f({\bf \theta}^{t,b}_m; {\bf z}_l, {\bf z}_{ul})$;
                \State $\mathcal L_{1} = \omega_1 \mathcal L_{CE}^s(({\bf \hat{p}}_l, {\bf \hat{p}}_{ul}), ({y}_l,{\hat y}_{ul})) +$
                 \Statex \quad\quad\quad\quad\quad\quad\quad $ \omega_2\mathcal L_{VAT}(({\bf r}_l, {\bf r}_{ul}))$;
            \Statex \quad\quad\quad\quad{\bf Backward propagation}:
                \State ${{\bf \theta}^{t,b+1}_m} \leftarrow Adam({\nabla}_{\theta_m}, \mathcal L_{1}, lr_m, \theta_m)$
            \EndIf
        \EndFor
    \EndFor
\end{algorithmic}
\end{algorithm}

\section{Experimental Setup and Results}
\label{sec5}

\subsection{Dataset Description}
We use the dataset proposed in paper \cite{Tu_2022} and \cite{Sankhe_2019} to evaluate our proposed MAT-based SS-SEI method. The former is a large-scale real-world radio signal dataset based on a special aeronautical monitoring system, ADS-B, and the latter is WiFi dataset collected from USRP X310 radios that emit IEEE 802.11a standards compliant frames. These dataset are suitable for evaluating the identification performance of SEI method. The number of categories of ADS-B dataset and WiFi dataset is $10$ and $16$, respectively. The length of each sample of ADS-B dataset and WiFi dataset is $4,800$ and $6,000$, respectively. The number of training samples of ADS-B dataset and WiFi datsset is $3,080$. The number of testing samples of ADS-B dataset and WiFi dataset is $1,000$ and $16,004$, respectively. We construct five semi-supervised scenarios and one fully supervised scenario, where the number of labeled training samples to the number of all training samples ratio is $\{5\%, 10\%, 20\%, 50\%, 100\%\}$, to evaluate the identification performance of the proposed SS-SEI method. In addition, 30\% of the training samples is used as the validating samples during the training process. These dataset are also available on https://github.com/lovelymimola/MAT-based-SS-SEI.

\subsection{Simulation Parameters}
We implement our approach in PyTorch \cite{Paszke_2017} (v1.10.2 with Python 3.6.13) by optimizing the parameters using Adam \cite{Kinga_2015} with
learning rates $lr_m = 0.001$, $lr_a$ of center loss and proxy anchor loss is $0.001$ and $0.05$, respectively, and other initial parameters of Adam. The scalars the objective functions (\ref{fun: joint loss2}) and (\ref{fun: joint loss1}) is automatically tuned by method \cite{liebei_2018}. The scaling factor of proxy-anchor loss is $\alpha = 32$ and margin of proxy-anchor loss is $\delta = 0.1$. The perturbation intensity of VAT is $\epsilon = 1.0$. We train the model for $300$ iterations and the batch size is $32$. Experiments are performed using NVIDIA GeForce RTX 3090 GPU.

\subsection{Comparative Methods}
In this paper, our SS-SEI method is compared with four latest SS-SEI or SS-SI methods, including DRCN \cite{WangY_2020}, SSRCNN \cite{Dong_2021}, Triple-GAN \cite{Gong_2019, Li_2017}, and SimMIM \cite{Xie_2022CVPR, Huang_2022}, where the SimMIM-based SEI method \cite{Huang_2022} uses the all training samples to pretrain the autoencoder and then uses the labeled training samples to finetune the encoder and classifier, and it is considered as a SS-SEI method for a comparison in this papaer. In addition, we compare the proposed MAT-based SS-SEI method with CVNN-based SEI method \cite{WangY_2021} and we only use the labeled training samples to train the CVNN-based SEI method. On the premise of not changing the core idea of these comparison methods, we use the same dataset with IQ format, data preprocessing, optimizer, learning rate and basic network structure for a fair comparison.

\subsection{Identification Performance: MAT VS. Comparative methods}

The labeled training dataset to training dataset ratio not only influences the identification performance of the SS-SEI method but also evaluates whether the SS-SEI methods have the ability to handle real environment. The identification performance of our MAT-based SS-SEI method and comparative methods under ADS-B and WiFi dataset are shown in Table \ref{Tab:Details of identification accuracy}. We observe the clear superiority of our SS-SEI method over comparative methods under ADS-B dataset and WiFi dataset. For the ADS-B dataset, when the labeled training dataset to training dataset ratio is $10\%$, the identification accuracy of comparative methods is more than $60\%$ but less than $80\%$, while our SS-SEI method can reach more than $80\%$. For the WiFi dataset, when the labeled training dataset to training dataset ratio is $10\%$, the identification accuracy of comparative methods is more than $20\%$ but less than $50\%$, while our SS-SEI method can reach more than $80\%$.

\begin{table*}[htbp]
  \caption{The identification accuracy of proposed MAT-based SS-SEI method under ADS-B dataset.}
  \begin{center}
  \begin{tabular}{cccccccccccc}
  \hline
  \multirow{2}{*}{\bf Methods}&\multicolumn{5}{c}{ADS-B}&&\multicolumn{5}{c}{WiFi}\\
  \cline{2-6} \cline{8-12}
  ~&{\bf 5\%}&{\bf 10\%}&{\bf 20\%}&{\bf 50\%}&{\bf 100\%}&&{\bf 5\%}&{\bf 10\%}&{\bf 20\%}&{\bf 50\%}&{\bf 100\%}\\
  \hline
  CVNN \cite{WangY_2021}&60.50\%&74.50\%&92.70\%&97.70\%&99.20\%&&20.47\%&28.64\%&69.78\%&97.14\%&99.46\%\\
  DRCN \cite{WangY_2020}&54.20\%&72.40\%&93.60\%&97.10\%&98.90\%&&21.94\%&47.51\%&76.18\%&98.99\%&99.64\%\\
  SSRCNN \cite{Dong_2021}&49.30\%&79.30\%&91.00\%&97.50\%&99.10\%&&19.33\%&38.09\%&99.25\%&99.75\%&99.76\%\\
  Triple-GAN \cite{Gong_2019}&45.10\%&61.10\%&90.90\%&97.10\%&99.10\%&&27.57\%&37.27\%&72.88\%&99.06\%&99.63\%\\
  SimMIM \cite{Huang_2022}&65.90\%&77.90\%&92.90\%&97.60\%&98.80\%&&31.71\%&49.59\%&75.76\%&96.01\%&99.41\%\\
  {\bf MAT-CL (Proposed)}&70.06\%&83.80\%&\bf 95.00\%&\bf 99.10\%&\bf 99.40\%&&27.26\%&\bf 80.70\%&\bf 99.76\%&\bf 99.79\%&\bf 99.79\%\\
  {\bf MAT-PA (Proposed)}&\bf 74.00\%&\bf 84.80\%&93.90\%&97.30\%&99.30\%&&\bf 28.82\%&54.96\%&98.18\%&99.77\%&99.77\%\\
  \hline
  \end{tabular}
  \end{center}
  \label{Tab:Details of identification accuracy}
\end{table*}

\subsection{Visualization of Semantic Features: MAT VS. Comparative methods}

\begin{figure*}[htbp]
  \centering
  \subfigure[CVNN\cite{WangY_2021}]{
   \label{CVNN.}
   \includegraphics[width=2.2 in] {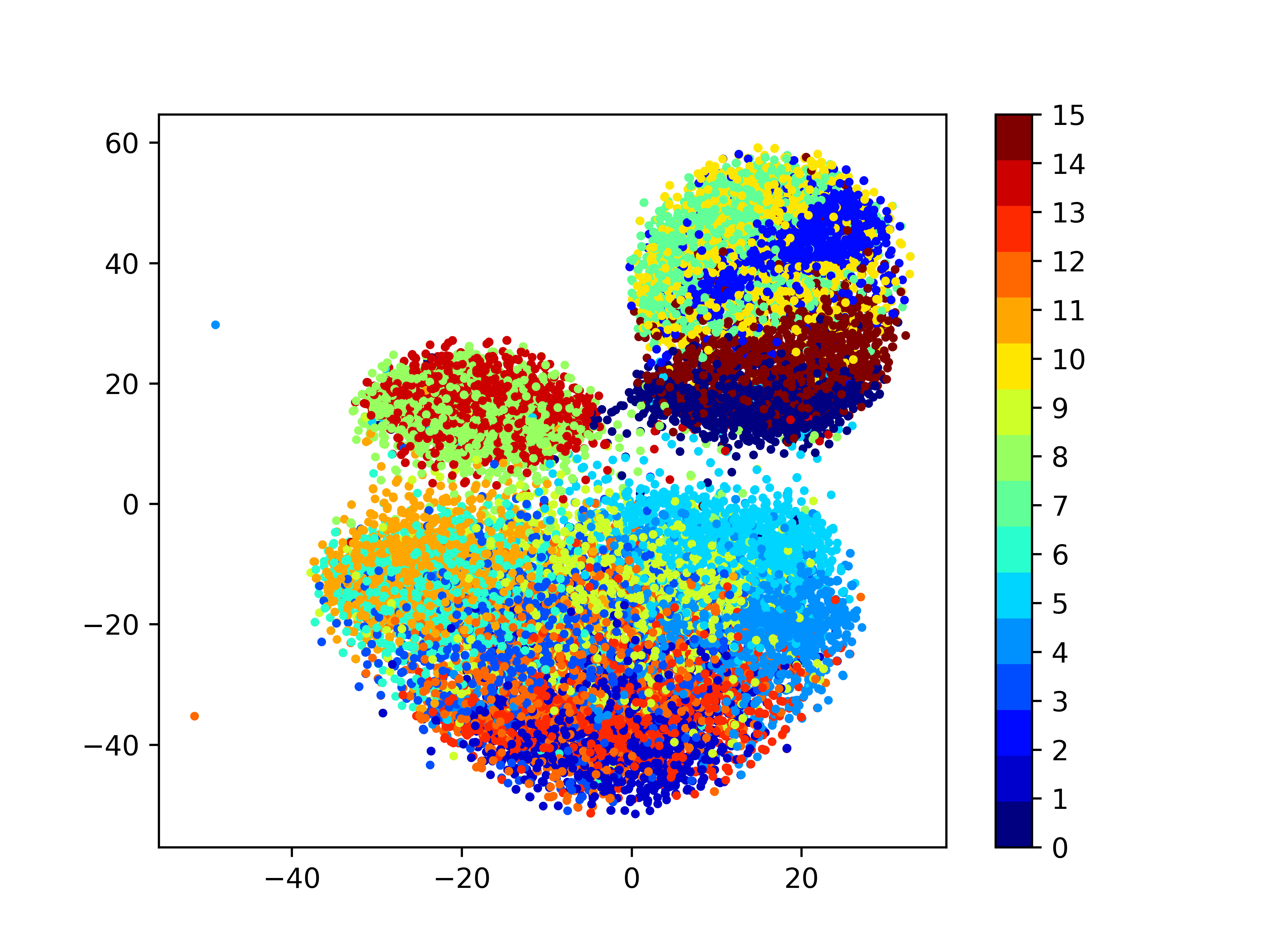}}
  \subfigure[DRCN \cite{WangY_2020}]{
   \label{DRCN.}
   \includegraphics[width=2.2 in] {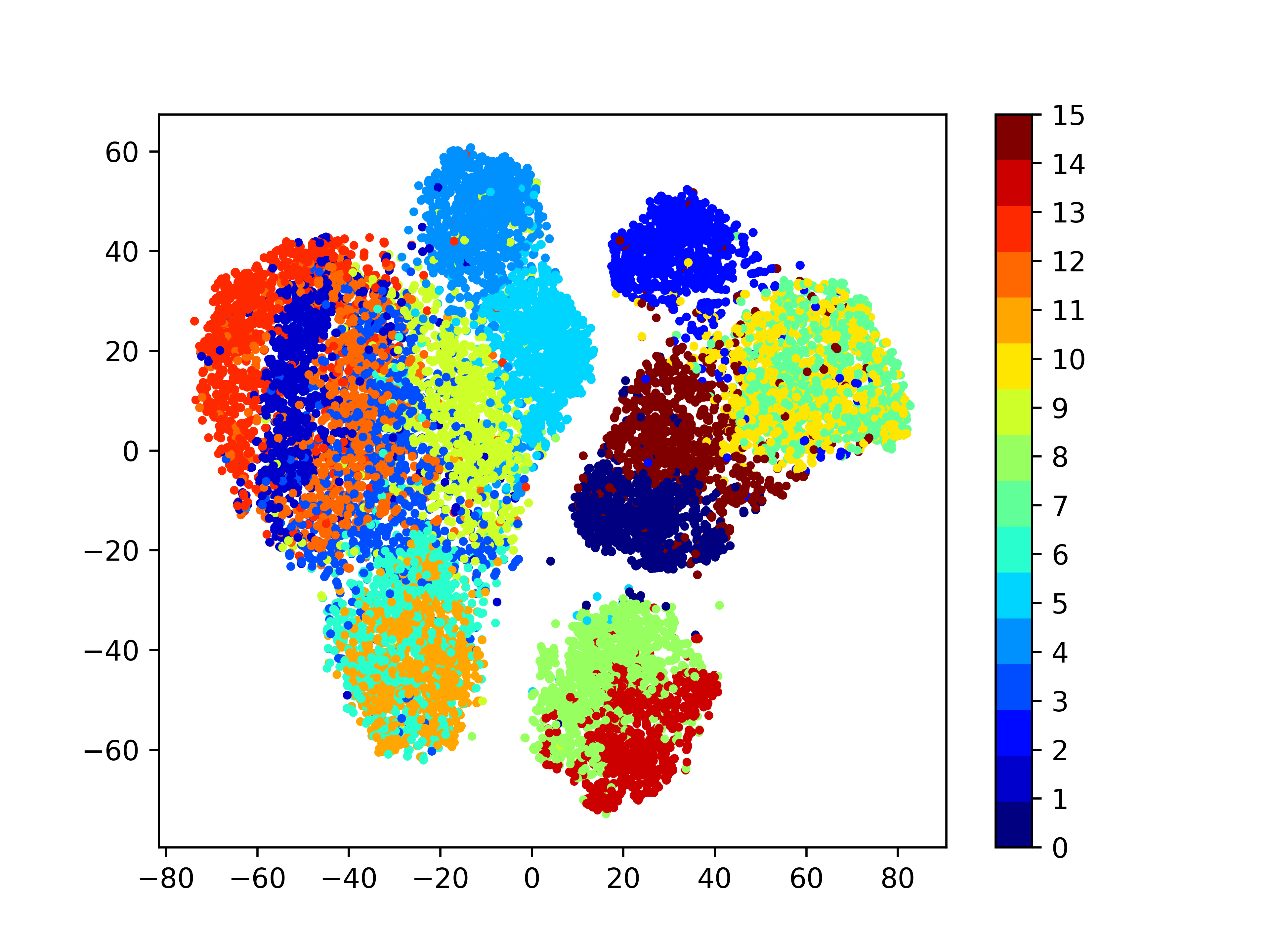}}
  \subfigure[SSRCNN \cite{Dong_2021}]{
   \label{SSRCNN.}
   \includegraphics[width=2.2 in] {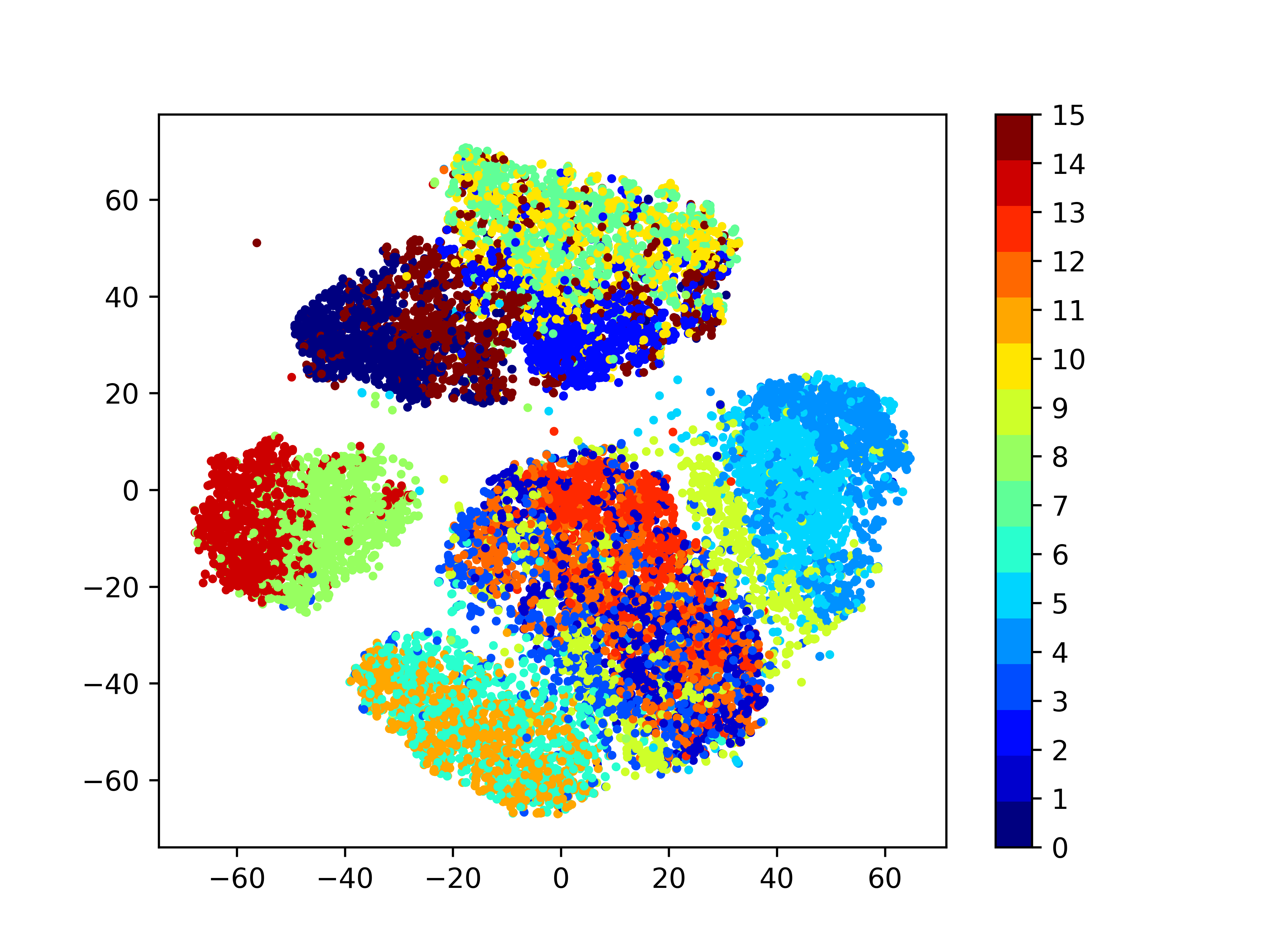}}
  \subfigure[Triple-GAN \cite{Gong_2019, Li_2017}]{
   \label{Triple-GAN.}
   \includegraphics[width=2.2 in] {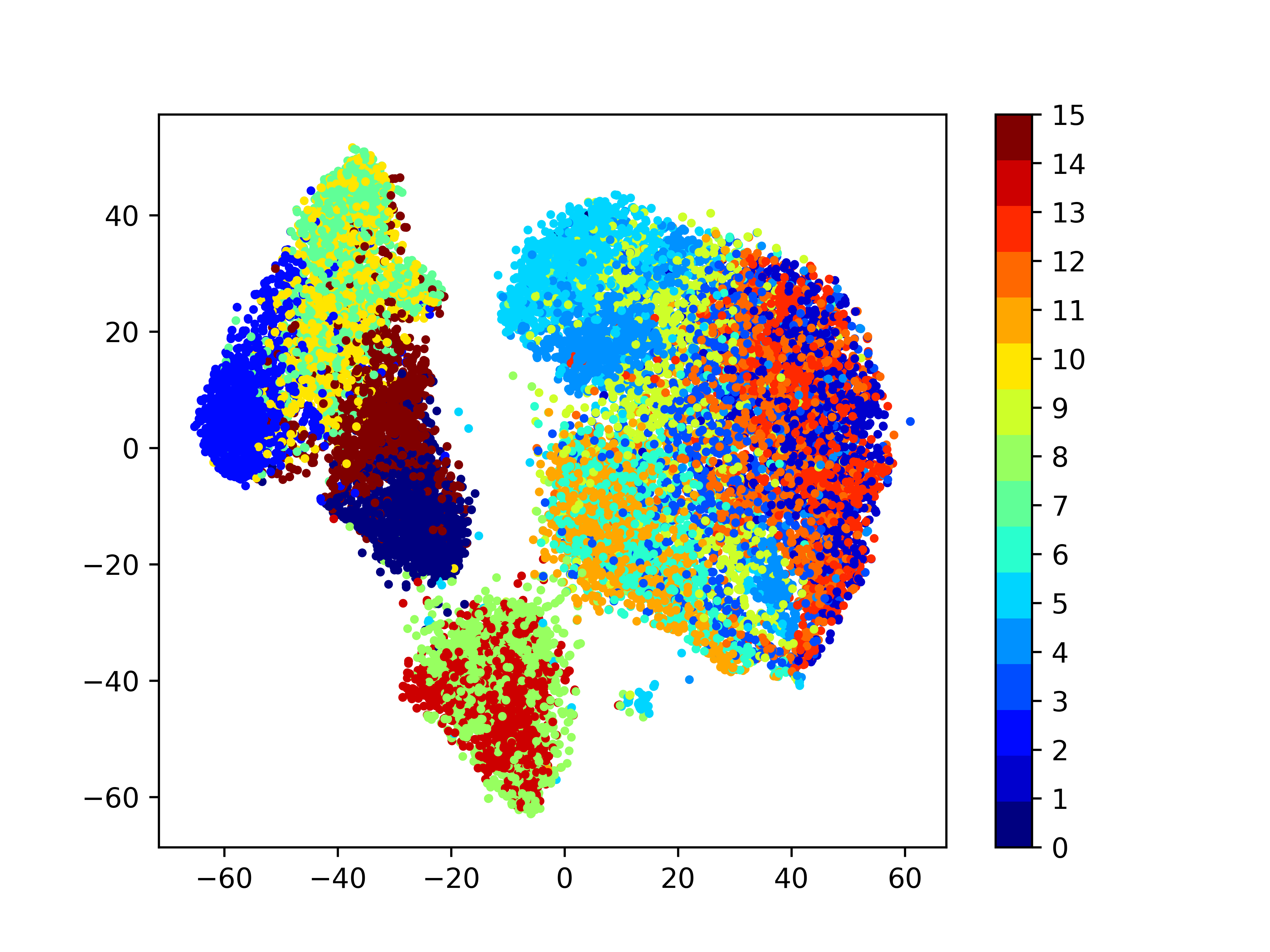}}
  \subfigure[SimMIM \cite{Huang_2022}]{
   \label{SimMIM.}
   \includegraphics[width=2.2 in] {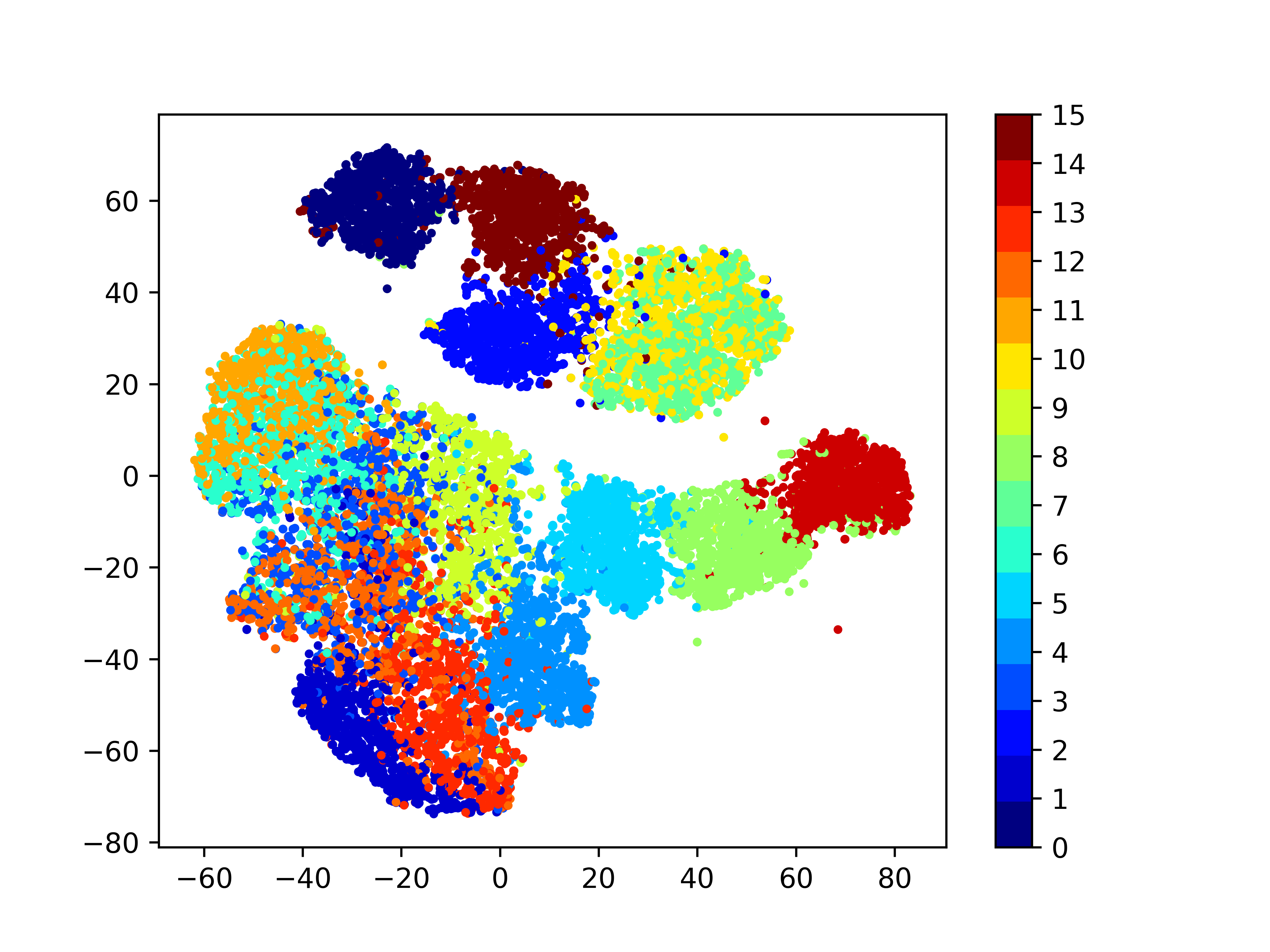}}
  \subfigure[MAT-CL (Proposed)]{
   \label{MAT.}
   \includegraphics[width=2.2 in] {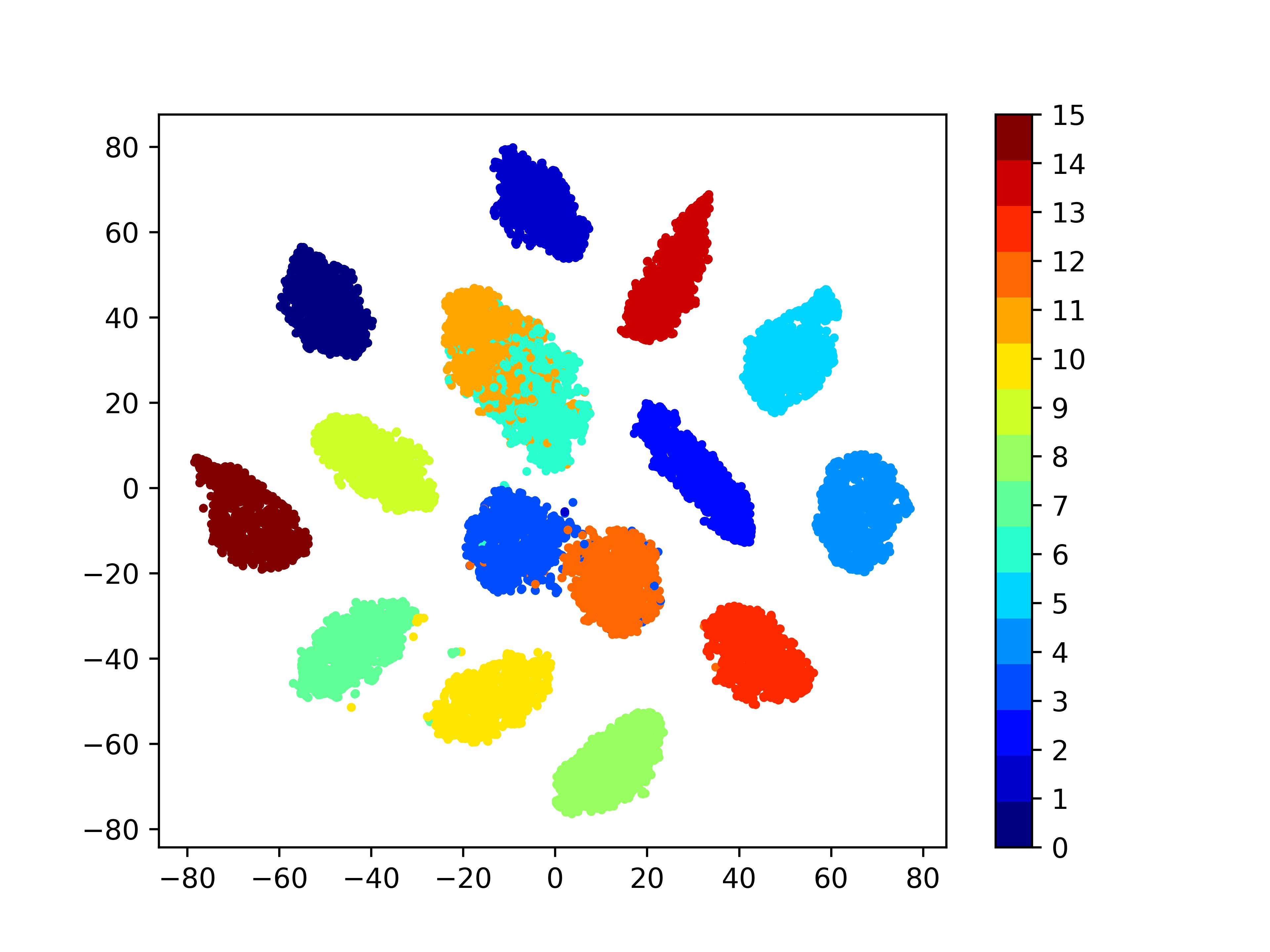}}
 \caption{Visualization of semantic features of different SEI-method when the dataset is WiFi and the labeled training dataset to training dataset ratio is 10\%, where the silhouette coefficient of CVNN, DRCN, SSRCNN, TripleGAN, SimMIM and MAT-CL is 0.0028, 0.0042, 0.0014, 0.015, 0.069 and 0.54, respectively.}
 \label{Visualization of Semantic Vectors}
\end{figure*}

A well-designed objective function is presented to extract the discriminative and generalized semantic features in this paper, where the objective function is alternatively regularized by SSML and VAT during training. The dimensionality of the extracted semantic features is reduced to two dimensions by t-distributed stochastic neighbor embedding (t-SNE) \cite{Maaten_2008} for visualization as shown in Fig. \ref{Visualization of Semantic Vectors}. We only show the visualization under WiFi dataset and the ratio is $10\%$ because of the limited space, and the visualization of other scenarios can be seen in our Github (https://github.com/lovelymimola/MAT-based-SS-SEI).

It can be observed that the semantic features of different categories extracted by CVNN staying apart roughly because CVNN is merely optimized by CE loss in a fully-supervised way and the data distribution information included in the limited labeled training dataset is insufficient. DRCN, SSRCNN, Triple-GAN and SimMIM not only use the data distribution information included in the limited labeled training dataset but also use the data distribution information included in the large unlabeled training dataset, and therefore the semantic features are more discriminative than that of the CVNN. We also observe the clear superiority of our MAT-based SS-SEI method over comparative methods in visualization of semantic features. Specifically, the semantic features with the inter-category dispersion and intra-category compactness are obtained by our proposed MAT-based SS-SEI method.

\subsection{Training Time: MAT VS. Comparative methods}
The proposed MAT is essentially a well-designed loss function and novel training strategy, which can be used in a variety of DNNs to identify different radio signals or emitters and the CVNN is used to verify the effectiveness of MAT in this paper. We analyze the average time per-iteration of training process as shown in Table \ref{Tab:Training Time}. It can be observed that the average time per-iteration of DRCN and TripleGAN is more than that of CVNN, SimMIM, MAT-CL and MAT-PA because the structure of DRCN contains not only encoder and classifier but also decoder, and TripleGAN contains not only encoder and classifier but also decoder, generator and discriminator. Therefore, DRCN-based SEI method and TripleGAN-based SEI method take extra time to train the decoder, generator and discriminator. The structure of neural network of SimMIM contains encoder, decoder and classifier, but the decoder is lightweight and the computational complexity does not increase sharply. Although the structure of neural network of MAT-CL and MAT-PA is same as the CVNN, the objective function of MAT-CL and MAT-PA is more complex than CVNN and thus the average time per-iteration of MAT-CL and MAT-PA is slightly more than that of CVNN.
\begin{table*}[hbtp]
  \caption{The average time per-iteration of training process of proposed MAT and comparative methods under WiFi dataset}
  \begin{center}
  \begin{tabular}{|c|c|c|c|c|c|c|c|}
  \hline
  Ratio&CVNN&DRCN&SSRCNN&TripleGAN&SimMIM&\bf MAT-CL (Proposed)&\bf MAT-PA (Proposed)\\
  \hline
  5\%&\bf 0.26s&3.80s&2.96s&3.14s&1.19s&1.93s&1.96s\\
  \hline
  10\%&\bf 0.39s&6.61s&4.59s&5.74s&1.29s&3.35s&3.36s\\
  \hline
  20\%&\bf 0.71s&7.90s&4.43s&10.56s&1.54s&4.30s&4.27s\\
  \hline
  50\%&\bf 1.68s&8.92s&3.91s&16.23s&2.15s&5.58s&5.39s\\
  \hline
  100\%&\bf 3.22s&12.37s&3.59s&32.43s&3.40s&6.88s&6.75s\\
  \hline
  \end{tabular}
  \end{center}
  \label{Tab:Training Time}
\end{table*}

\subsection{Ablation Analysis of Proposed MAT}
We include an extensive ablation study to tease apart the importance of the different components of MAT. In addition, the ablation analysis which is often ignored in SS-SEI methods, MAT with supervised algorithm that uses only labeled data, is considered in this paper. Ablation details are shown in the Table \ref{Tab:Details of Ablation Analysis}. The identification performance with different ablation analysis are shown as Table \ref{Tab:Identification performance of Ablation Analysis}. It can be observed that all of factors are crucial to MAT-based SEI method's success, and the identification performance of MAT will decrease if any factor is ablated. It also can be observed that the identification performance of MAT-* w/o UTD is better than that of MAT-* under some scenarios because the MAT is sensitive to the amount of labeled and unlabeled data which is the shortcoming of most SSL method \cite{Oliver_2018}.

\begin{table}[htbp]
  \caption{Details of ablation analysis.}
  \begin{center}
  \begin{tabular}{ccccc}
  \hline
  {\bf Components}&{\bf SS-CE}&{\bf SSML}&{\bf VAT}&{\bf UTD}\\
  \hline
  MAT-* w/o SSML &$\checkmark$&&$\checkmark$&$\checkmark$\\
  MAT-* w/o VAT &$\checkmark$&$\checkmark$&&$\checkmark$\\
  MAT-* w/o UTD &$\checkmark$&$\checkmark$&$\checkmark$&\\
  \bf MAT-* (Proposed)&$\checkmark$&$\checkmark$&$\checkmark$&$\checkmark$\\
  \hline
  \multicolumn{5}{l}{\small UTD denotes the unlabeled training samples}\\
  \multicolumn{5}{l}{\small and * means the SSML is SS-CL of SS-PA}\\
  \multicolumn{5}{l}{\small and w/o is an abbreviation for without.}\\
  \end{tabular}
  \end{center}
  \label{Tab:Details of Ablation Analysis}
\end{table}

\begin{table}[htbp]
  \caption{Ablation analysis of proposed MAT.}
  \begin{center}

  \subtable[Ablation analysis of proposed MAT-PA under ADS-B and WiFi dataset]{
  \begin{tabular}{cccccc}
  \hline
  \multirow{2}{*}{\bf Methods}&\multicolumn{2}{c}{ADS-B}&&\multicolumn{2}{c}{WiFi}\\
  \cline{2-3} \cline{5-6}
  ~&{\bf 5\%}&{\bf 50\%}&&\bf 5\%&\bf 10\%\\
  \cline{1-3} \cline{5-6}
  MAT-PA w/o SSML &69.60\%&97.40\%&&24.32\%&99.14\%\\
  MAT-PA w/o VAT &61.00\%&96.80\%&&31.27\%&84.77\%\\
  MAT-PA w/o UTD &71.40\%&\bf 98.70\%&&\bf 32.15\%&99.15\%\\
  \bf MAT-PA (Proposed) &\bf 74.00\%&97.30\%&&28.82\%&\bf 99.77\%\\
  \hline
  \end{tabular}}

  \subtable[Ablation analysis of proposed MAT-CL under ADS-B and WiFi dataset]{
  \begin{tabular}{cccccc}
  \hline
  \multirow{2}{*}{\bf Methods}&\multicolumn{2}{c}{ADS-B}&&\multicolumn{2}{c}{WiFi}\\
  \cline{2-3} \cline{5-6}
  ~&{\bf 5\%}&{\bf 50\%}&&\bf 5\%&\bf 10\%\\
  \cline{1-3} \cline{5-6}
  MAT-CL w/o SSML &69.60\%&97.40\%&&31.27\%&84.77\%\\
  MAT-CL w/o VAT &57.60\%&97.30\%&&24.52\%&97.57\%\\
  MAT-CL w/o UTD &65.80\%&97.90\%&&\bf 34.92\%&99.71\%\\
  \bf MAT-CL (Proposed) &\bf 70.06\%&\bf 99.10\%&&27.26\%&\bf 99.79\%\\
  \hline
  \end{tabular}}

  \end{center}
  \label{Tab:Identification performance of Ablation Analysis}
\end{table}


\subsection{Alternating Optimization VS. Simultaneous Optimization}
\begin{figure}[htbp]
  \centering

  \subfigure[]{
   \label{MAT-PA}
   \includegraphics[width=3.2 in] {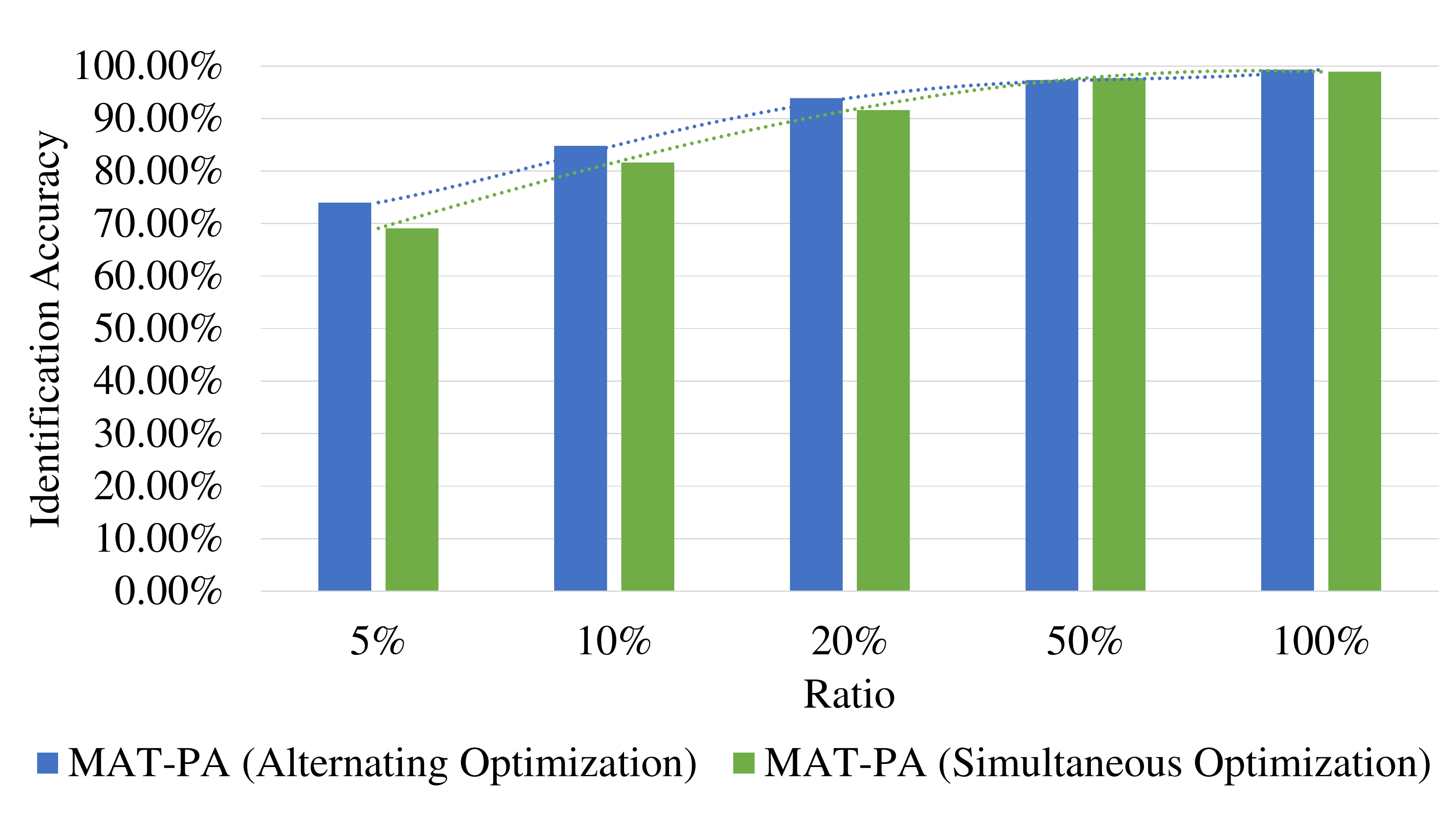}}

  \subfigure[]{
   \label{MAT-CL}
   \includegraphics[width=3.2 in] {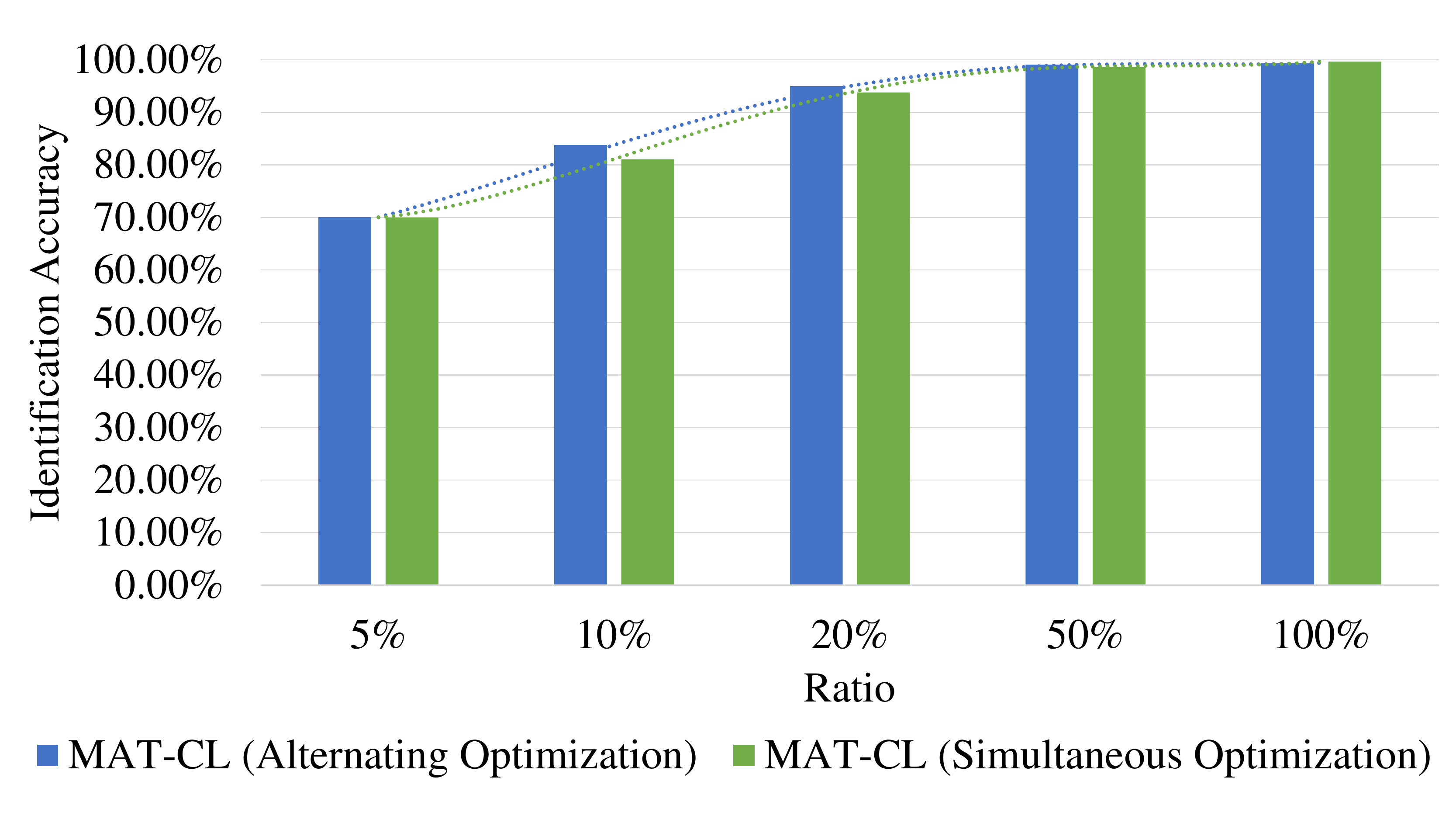}}

 \caption{The identification performance of two optimization approaches on ADS-B dataset, where the Ratio is the number of labeled training samples to the number of all training samples ratio.}
 \label{The identification performance of two optimization approaches on ADS-B dataset}
\end{figure}

\begin{figure}[htbp]
  \centering

  \subfigure[The training loss of MAT-PA]{
   \label{MAT-PA}
   \includegraphics[width=2.8 in] {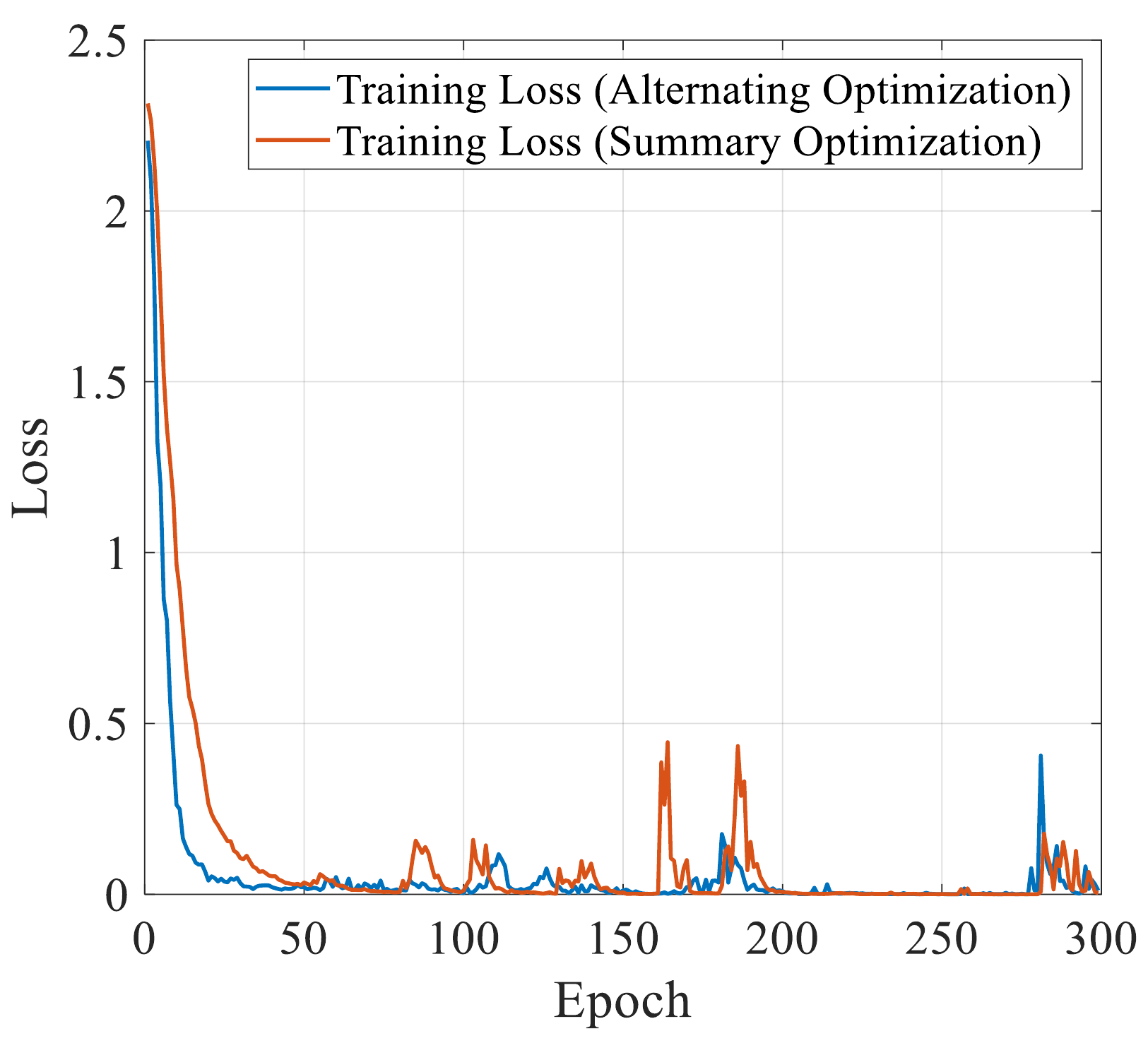}}

  \subfigure[The training loss of MAT-CL]{
   \label{MAT-CL}
   \includegraphics[width=2.8 in] {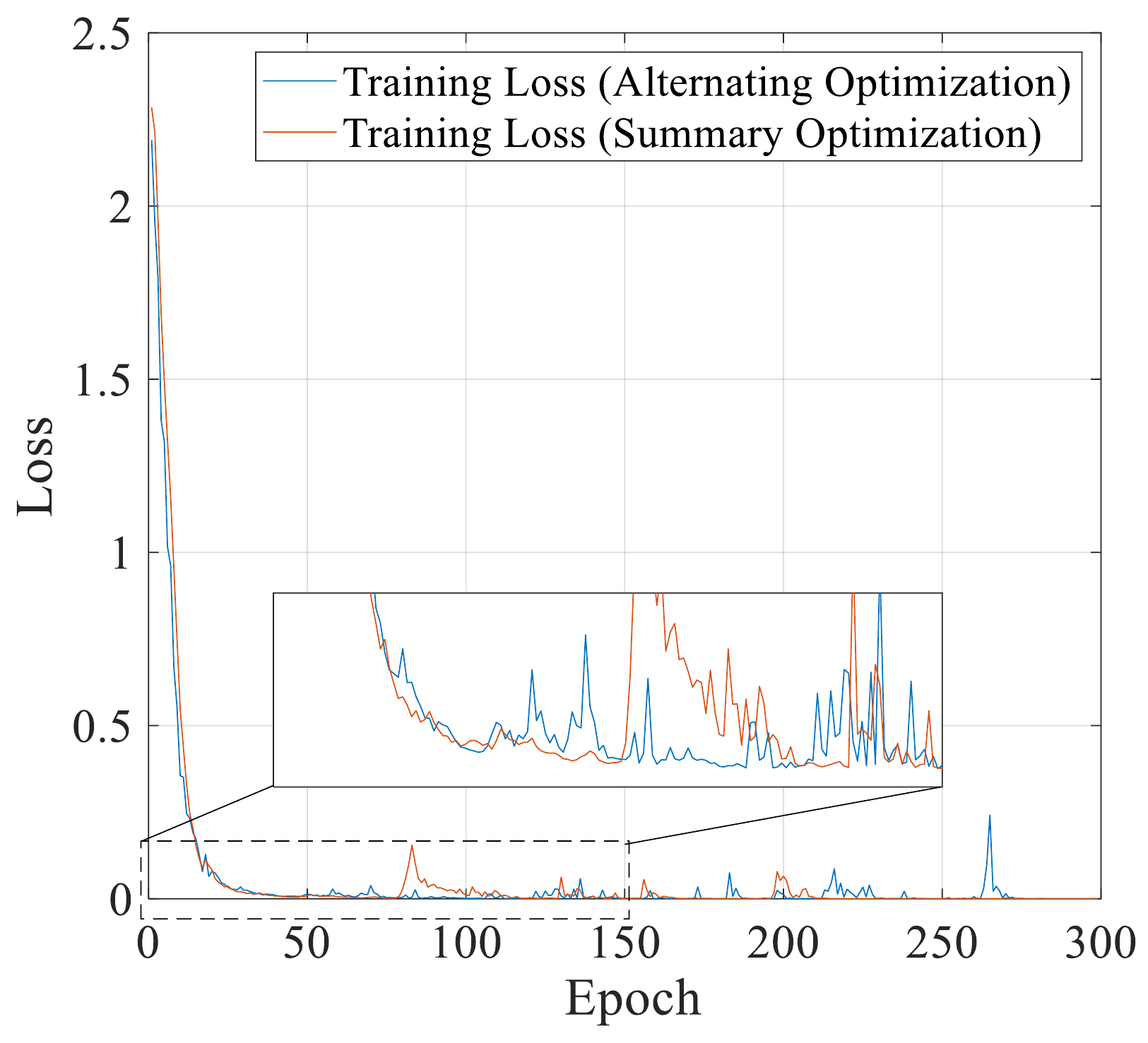}}

 \caption{The training loss of two optimization approaches under ADS-B dataset and the number of labeled training samples to the number of all training samples is 10\%.}
 \label{The training loss of two optimization approaches}
\end{figure}

In this paper, the objective function is alternatively regularized by SSML and VAT that is extremely different from standard optimization termed simultaneous optimization, i.e.,
\begin{equation}
\begin{split}
\label{fun: joint loss2}
{\mathcal L} = {\omega}_1 {\mathcal L}_{CE}^s + {\omega}_2 {\mathcal L}_{VAT} + {\omega}_3 {\mathcal L}_{SSML}.
\end{split}
\end{equation}
The identification performance of two optimization approaches on ADS-B dataset is shown in Figure \ref{The identification performance of two optimization approaches on ADS-B dataset}. It can be observed the clear superiority of identification performance of alternating optimization over simultaneous optimization, and the gaps of identification performance between alternating optimization and simultaneous optimization are $[-0.30\%, 2.70\%]$ and $[-0.40\%, 4.90\%]$ in Fig. \ref{MAT-PA} and Fig. \ref{MAT-CL}, respectively. The training loss of two optimization approaches under ADS-B dataset and the number of labeled training samples to the number of all training samples ratio is $10\%$ is shown in Fig. \ref{The training loss of two optimization approaches}. It can be observed that the convergence rate of alternating optimization is faster than that of simultaneous optimization. The advantage of alternating optimization is the higher identification performance and faster convergence rate.

\subsection{SSML VS. ML}
\begin{figure}[htbp]
  \centering
  \subfigure[MAT with CL]{
   \label{MAT with CL.}
   \includegraphics[width=3.2 in] {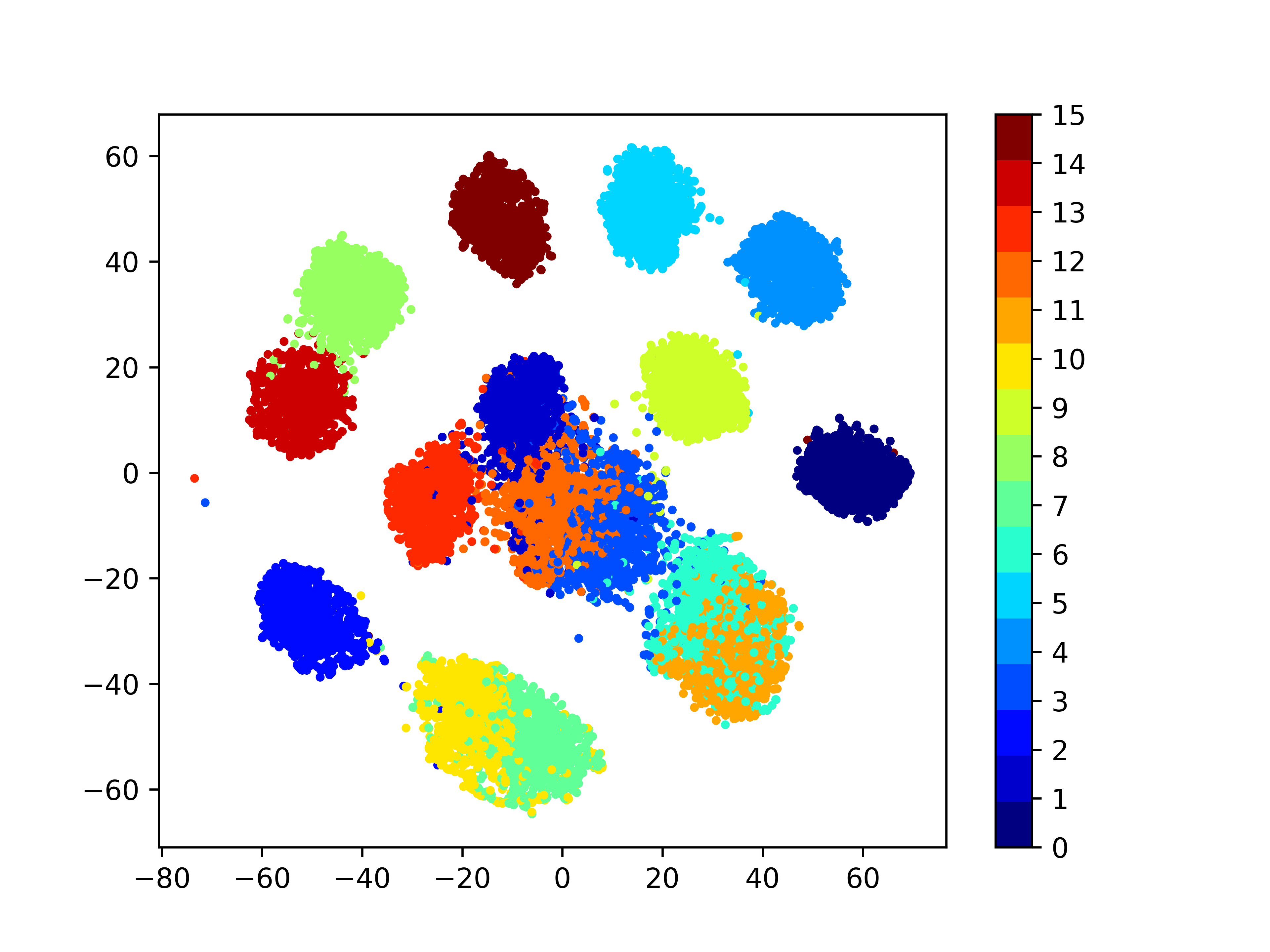}}
  \subfigure[MAT with SS-CL]{
   \label{MAT with SS-CL.}
   \includegraphics[width=3.2 in] {fig/CNN_MAT_n_classes_16_10label_90unlabel_improved.png}}
 \caption{Visualization of semantic features of MAT with CL or SS-CL when the dataset is WiFi and the number of labeled training samples to the number of all training samples ratio is 10\%, where the silhouette coefficient of MAT with CL and MAT with SS-CL is 0.13 and 0.54, respectively.}
 \label{Visualization of Semantic Vectors of MAT with ML or SSML}
\end{figure}

The identification performance of ML and SSML is shown as Table \ref{Tab:The identification performance of standard metric and semi-supervised metric}, where the $(\uparrow)$ means the identification performance of SSML is better than that of ML, and the $(\downarrow)$ means the identification performance of SSML is worse than that of ML, and the $(-)$ means the identification performance of SSML is same as the that of ML, and the ratio is the number of labeled samples to the number of all samples. The visualization of semantic features extracted by MAT with CL and MAT with SS-CL when ratio is $10\%$ are shown in Fig \ref{MAT with CL.} and Fig \ref{MAT with SS-CL.}, respectively. It can be observed that the proposed trick can improve the identification performance and increase the inter-category dispersion and intra-category compactness of the extracted semantic features. In addition, the tolerable loss (i.e., $0.50\%$ - $2.70\%$) of identification performance is brought by SSML due to the sensitivity of SSL to the amount of labeled and unlabeled data.

\begin{table*}[hbpt]
  \caption{The identification performance of ML and SSML.}
  \begin{center}
  \begin{tabular}{|c|c|c|c|c|c|c|c|c|}
  \hline
  \multirow{2}{*}{\bf Ratio}&\multicolumn{2}{|c|}{MAT-CL (ADS-B)}&\multicolumn{2}{|c|}{MAT-PA (ADS-B)}&\multicolumn{2}{|c|}{MAT-CL (WiFi)}&\multicolumn{2}{|c|}{MAT-PA (WiFi)}\\
  \cline{2-9}
  ~&ML&SSML&ML&SSML&ML&SSML&ML&SSML\\
  \hline
  5\%&70.50\%&70.50\% $(-)$&72.00\%&74.00\% $(\uparrow)$&29.33\%&27.26\% $(\downarrow)$&28.82\%&28.82\% $(-)$\\
  \hline
  10\%&86.50\%&83.80\% $(\downarrow)$&84.80\%&84.80\% $(-)$&57.43\%&80.70\% $(\uparrow)$&54.96\%&54.96\% $(-)$\\
  \hline
  20\%&95.00\%&95.00\% $(-)$&93.60\%&93.90\% $(\uparrow)$&99.76\%&99.76\% $(-)$&99.70\%&98.18\% $(\downarrow)$\\
  \hline
  50\%&98.60\%&99.10\% $(\uparrow)$&97.80\%&97.30\% $(\downarrow)$&99.79\%&99.79\% $(-)$&99.76\%&99.77\% $(\uparrow)$\\
  \hline
  100\%&99.40\%&99.40\% $(-)$&99.30\%&99.30\% $(-)$&99.79\%&99.79\% $(-)$&99.77\%&99.77\% $(-)$\\
  \hline

  \hline
  \end{tabular}
  \end{center}
  \label{Tab:The identification performance of standard metric and semi-supervised metric}
\end{table*}

\section{Conclusion}
\label{sec6}
In this paper, we proposed a SS-SEI method using MAT. Specifically, pseudo labels are innovatively introduced into ML and the SSML was proposed and used to extract the discriminative semantic features. VAT was used to extract the generalized semantic features. More Specifically, an object function (i.e., the cross-entropy loss regularized by SSML or VAT) and an alternating optimization way were designed to achieve a SS-SEI method. The proposed MAT-based SS-SEI method was evaluated on an open source large-scale real-word ADS-B dataset and WiFi dataset and was compared with four latest SS-SEI methods. The simulation results showed that the proposed MAT-based SS-SEI method achieved the state-of-the-art identification performance. When the labeled training dataset to training dataset ratio was $10\%$, the identification accuracy of MAT-CL was $83.80\%$ under ADS-B dataset and $80.70\%$ under WiFi dataset. Evaluating the identification performance of proposed MAT-based SS-SEI method on multiple SEI datatset and achieving open-set SEI identification based on MAT is our future work.

\end{document}